\documentclass{aastex}
\usepackage{spr-astr-addons}
\usepackage{natbib}
\usepackage{url}
\pdfoutput=1

\RequirePackage{color}

\begin{document}

\title{RR Lyrae stars and the horizontal branch of NGC 5904 (M5)}
\slugcomment{Not to appear in Nonlearned J., 45.}
\shorttitle{The RR Lyrae stars in NGC 5904 (M5)}
\shortauthors{Arellano Ferro et al.}

\author{A. Arellano Ferro\altaffilmark{1}}  \author{A. Luna\altaffilmark{1}}
\affil{Instituto de Astronom\'ia, Universidad Nacional Aut\'onoma de M\'exico.
Ciudad Universitaria CP 04510, Mexico:(armando@astro.unam.mx)} \author{D. M.
Bramich\altaffilmark{2}}
\affil{Qatar Environment and Energy Research Institute (QEERI), HBKU, Qatar Foundation,  Doha, Qatar: (dan.bramich@hotmail.co.uk)}  \author{Sunetra
Giridhar\altaffilmark{3}}
\affil{Indian Institute of Astrophysics, Koramangala 560034, Bangalore, India:
(giridhar@iiap.res.in)}
\author{J. A. Ahumada\altaffilmark{4}}
\affil{Observatorio Astron\'omico, Universidad Nacional de
C\'ordoba, Laprida 854, 5000 C\'ordoba, Argentina: (javier@oac.uncor.edu)}
\author{S. Muneer\altaffilmark{3}}
\affil{Indian Institute of Astrophysics, Koramangala 560034, Bangalore, India:
(muneer@iiap.res.in)}

\begin{abstract}
We report the distance and [Fe/H] value for the globular cluster NGC 5904 (M5)
derived from
the
Fourier decomposition of the light curves of selected RRab and RRc stars. The aim in
doing this was
to bring these parameters into the homogeneous scales established by our previous work
on numerous other
globular clusters, allowing a direct comparison of the horizontal branch luminosity in
clusters with a wide range of metallicities.
Our CCD photometry of the large variable star population of this cluster is
used to discuss light curve peculiarities,
like Blazhko modulations, on an individual basis. New Blazhko variables are reported. 

From the RRab stars we found [Fe/H]$_{\rm UVES} = -1.335 \pm
0.003{\rm(statistical)} \pm 0.110{\rm(systematic)}$, and 
a distance of $7.6\pm 0.2$ kpc, and from the RRc stars we found [Fe/H]$_{\rm
UVES}$ = $-1.39 \pm 0.03{\rm(statistical)} \pm 0.12{\rm(systematic)}$ and a distance
of $7.5 \pm 0.3$ kpc. The results for RRab and RRc stars should be
considered independent since they come from different calibrations and zero points. 
Absolute magnitudes, radii and masses are also reported for individual RR Lyrae
stars. 
The distance to the cluster was also calculated  by alternative methods like the
Period-Luminosity
relation of SX Phe and the luminosity of the stars at the tip of the red giant
branch, and we obtained
the results $7.7 \pm 0.4$ and $7.2-7.5$ kpc respectively.

The distribution of RR Lyrae stars in the instability strip is discussed  and
compared with other clusters in connection with the Oosterhoff and horizontal branch
type. The Oosterhoff type II clusters systematically show a RRab-RRc segregation about
the instability strip
first-overtone red edge, while the Oosterhoff type I clusters may or may not
display this feature.
A group of RR Lyrae stars is identified in an advanced evolutionary stage, and two of
them are likely binaries with unseen companions.
\end{abstract}

\keywords{globular clusters: individual (NGC~5904) -- stars:variables:
RR Lyrae, SX Phe, SR }


\section{Introduction}
The globular cluster NGC~5904 (M5, or
C1645+476 in the IAU nomenclature) ($\alpha = 15^{\mbox{\scriptsize h}}
18^{\mbox{\scriptsize m}} 33.2^{\mbox{\scriptsize s}}$, $\delta = +02\degr 04\arcmin
51.7\arcsec$, J2000; $l = 3.85\degr$, $b = +46.80\degr$) is among the closest
globular clusters (GCs) to the Sun and hence it is very bright. Its horizontal branch
is at
about $V \sim$15 mag. Being a nearby cluster, it is affected by very
little reddening, $E(B-V)=0.03$ mag (Harris 1996).

M5 has a very rich population of variable stars and no doubt its proximity has
contributed to
the very early discovery of numerous variables. The first 46 variables were
discovered 
by Solon J. Bailey in the last decade of the XIX century, on photographs taken with
the 13-inch Boyden Telescope at Arequipa, Peru, and they were announced 
by Pickering (1896a). Periods of some of these variables were calculated by Pickering 
(1896b) and Barnard (1898). Bailey himself reported the periods for 63 of the nearly 90
variables then known (Bailey \& Leland 1899) and in 1902 he listed X,Y
positions for 92 variables
(V1-V92, his table XXIV) and offered an identification chart (his Fig. 2, Bailey
1902).
Despite the richness of the cluster in variable stars, the next batch of discoveries only
happened about forty years later when Oosterhoff (1941) found V93-V103. Yet another
46 years later, Kadla et al. (1987) found the variables V104-V114. A year later
variables V115-V131 were announced by Kravtsov (1988) while V132-V133 were found by
Kravtsov (1991). Intense CCD photometry of the cluster led to the discovery of 35
more variables between 1996 and 2000; V134-V141 were announced by Sandquist et al.
(1996) although V134 and V135 were the already known variables V129 and V36
respectively.
In some cases the names of the new variables given in the original papers were 
modified by Caputo et al. (1999) or in the Catalogue of Variable Stars in Globular
Clusters (CVSGC; Clement et al.2001) to yield a consistent list of variable names.
Thus V142-143 were found by Brocato et al. (1996), V144-V148 by Reid
(1996), V149-V154 by Yan \& Reid (1996), V155-159 by Drissen \& Shara (1998),
V160-V163 by Olech et al. (1999) and V164-V168 by Kaluzny et al. (1999). V169 was noted by
Rees (1993) and confirmed by Kaluzny et al. (2000). The last batch of variables,
V170-V181, one SX Phe (V170) and 11 semi-regular late-type (SRA) variables,
was recently announced by
Arellano Ferro et al. (2015a). This makes nearly 120 years of variable star
discoveries in this cluster.

As in most of our recent papers we have employed the
{\tt DanDIA} \footnote{{\tt DanDIA} is built from the DanIDL library of IDL routines
available at http://www.danidl.co.uk} implementation of difference image analysis
(DIA) (Bramich 2008; Bramich et al. 2013)
to extract high-precision photometry for all of the point
sources in the field of M5. We collected 6890 light cuves in the $V$ and $I$
bandpasses
with the aim of building up a colour-magnitude diagram (CMD) and discussing the
horizontal branch (HB) structure as compared to other
Oosterhoff type I (OoI) and Oosterhoff type II (OoII) clusters.\footnote{Oosterhoff
(1939) noticed that GCs can be distinguished by the period
distribution of their RR  Lyrae stars; the mean period of fundamental pulsators or
RRab stars is $~\sim 0.55$ days in the type I or OoI and $\sim 0.65$ days in the type
II or OoII. Also the percentage of first overtone pulsators or RRc stars is higher in
OoI
clusters.} 
We also Fourier decompose the light curves of the RR Lyrae stars (RRL) to
calculate
their metallicity and luminosity in order to provide independent and homogeneous estimates
of the cluster mean metallicity and distance.
 
The scheme of the paper is as follows: In $\S$ 2 we describe the observations, data
reduction and calibration to the standard system. In $\S$ 3 the periods and phased
light curves of RRL stars are displayed and the Fourier light
curve decomposition of stable RRL is described. The corresponding
individual values of [Fe/H] and $M_V$ are reported. $\S$ 4 deals with the
discussion of
the distribution of RRL in the HB. In $\S$ 5 the metallicity and distance of
the parent cluster
are inferred from the RRL and independent distance calculations from the P-L
relation of SX Phe stars, and the luminosity of the tip of the red giant branch
(TRGB).
In this section we also give comments on peculiar stars.
Finally, in $\S$ 6 we summarize our conclusions.

\begin{table}[t]
\footnotesize
\caption{The distribution of observations of M5.
Columns $N_{V}$ and $N_{I}$ give the number of images taken with the $V$ and $I$
filters respectively. Columns $t_{V}$ and $t_{I}$ provide the exposure time,
or range of exposure times. The 
average seeing is listed in the last column.}
\centering
\begin{tabular}{@{}lccccc}
\hline
Date  &  $N_{V}$ & $t_{V}$ (s) & $N_{I}$ &$t_{I}$ (s)&Avg seeing (") \\
\hline
20120229 & 38 & 50-90 & 38 & 18-30 & 2.6\\
20120302 & 61 & 25-150 & 60 & 8-60 & 1.9 \\
20120411 & 26 & 25-600 & 25 & 8-300 & 2.2\\
20120428 & 28 & 20-250 & 29 & 10-160 & 2.2\\
20120513 & 68 & 14-45 & 65 & 5-12 & 1.7\\
20120514 & 1 & 70 & 5 & 10-80 & 1.6\\
20120515 & 37 & 20-60 & 33 & 7-30 & 1.9\\
20130119 & 5 & 30-90 & 3 & 15-30 & 2.0\\
20130730 & 21 & 18-30 & 20 & 3-10 & 1.4\\
20140408 & 66 & 10-12 & 68 & 4-5 & 1.8\\
20140409 & 34 & 10 & 38 & 4 & 1.7\\

\hline
Total:   & 385    &        &  384  &           &\\
\hline
\end{tabular}
\label{tab:observations}
\end{table}

\section{Observations and Reductions}
\label{sec:Observations}

\begin{figure} 
\includegraphics[width=8.0cm,height=8.0cm]{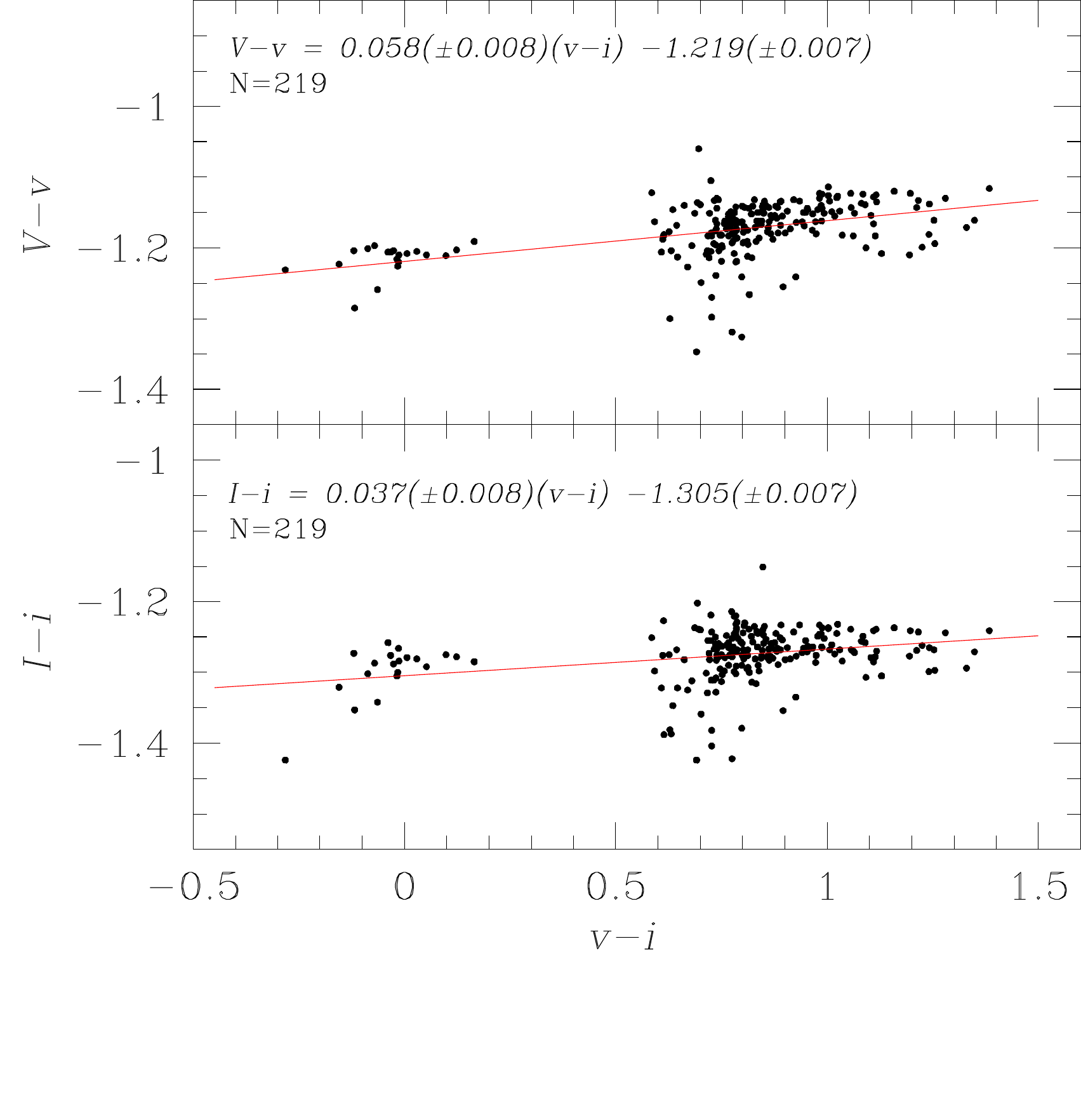}
\caption{Transformation relations in the $V$ and $I$ band-passes between the
instrumental and the standard
photometric systems using a set of standard stars in the field of M5 (Stetson 2000).}
    \label{transV}
\end{figure}

\begin{figure}
\includegraphics[width=8.0cm,height=8.0cm]{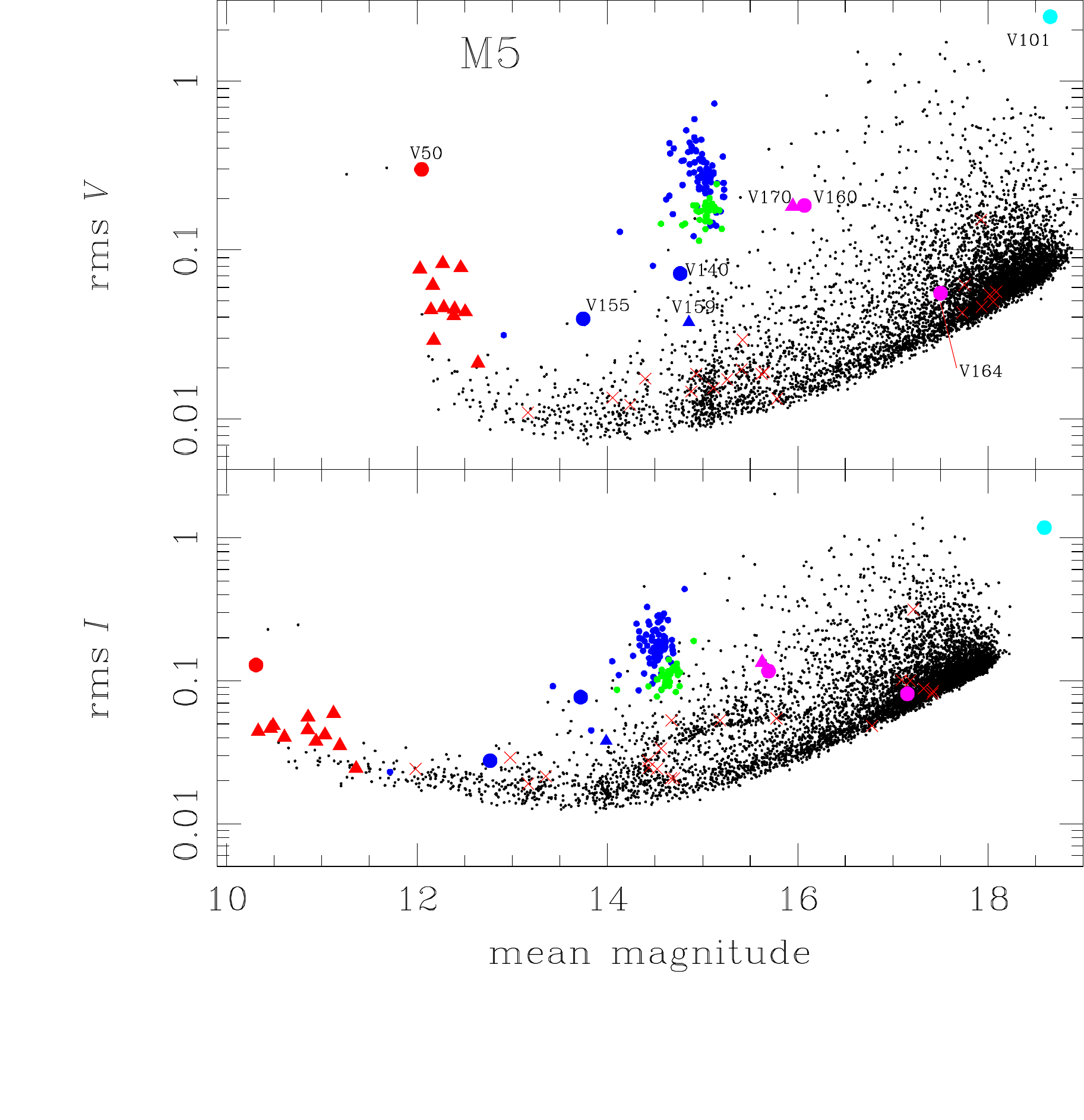}
\caption{The rms magnitude deviations as a function of
the mean magnitudes $V$ and $I$. Small blue and green circles represent RRab and RRc
stars respectively. Other symbols are: bigger and labelled blue filled symbols for
eclipsing binaries; red filled symbols for semi-regular late-type variables (SRA);
purple for SX Phe stars; red crosses are stars listed in the CVSGC as variables
but whose variability has not been confirmed by Arellano Ferro et al. (2015a). The 
U~Gem variable V101 is shown as a turquoise symbol. Triangles are used for all
variables
discovered by Arellano Ferro et al. (2015a) where a detailed discussion on all of them
can be found.}
\label{rmsCMD}
\end{figure}

\subsection{Observations}

The observations were performed on 11 nights between February 29, 2012 and  April 09,
2014 with the 2.0-m telescope at the Indian Astronomical Observatory (IAO), Hanle,
India,
located at 4500~m above sea level. A total of 385 and 384 images were obtained  in
the Johnson-Kron-Cousins $V$ and $I$ filters, respectively. The detector was a
Thompson CCD of 2048$\times$2048
pixels with a scale of 0.296 arcsec/pix, translating to a field of view (FoV) of
approximately
10.1$\times$10.1~arcmin$^2$. 

The log of observations is given in Table \ref{tab:observations} where the dates,
number of frames, exposure times and average nightly seeing are recorded.

\subsection{Difference Image Analysis}
We employed the technique of difference image analysis (DIA) to extract high-precision
photometry for all of the point sources in the images of M5 and we used the 
{\tt DanDIA}
pipeline for the data reduction process (Bramich 2008; Bramich et al. 2013). We constructed one
reference image for the $V$ filter and another for the $I$ filter by
stacking the best-quality images in our collection; then we
created sequences of difference images in each filter by subtracting the respective
convolved reference image from the rest of the collection. Differential fluxes for each
star detected in the reference
image were then measured on each difference image. Light curves for
each star were constructed by calculating the total fluxes which in turn were
converted
into instrumental magnitudes. The above procedure and its caveats have been
described
in detail in Bramich et al. (2011), so that for brevity we do not repeat them here
and refer the interested reader to that paper for further details.

\begin{table*}
\footnotesize
\caption{Time-series $V$ and $I$ photometry for all the variables in our 
field of view. The standard $M_{\mbox{\scriptsize std}}$ and
instrumental $m_{\mbox{\scriptsize ins}}$ magnitudes are listed in columns 4 and 5,
respectively, corresponding to the variable star in column 1. Filter and epoch of
mid-exposure are listed in columns 2 and 3, respectively. The uncertainty on
$m_{\mbox{\scriptsize ins}}$ is listed in column 6, which also corresponds to the
uncertainty on $M_{\mbox{\scriptsize std}}$. For completeness, we also list the
quantities
$f_{\mbox{\scriptsize ref}}$, $f_{\mbox{\scriptsize diff}}$ and $p$, 
in columns 7, 9 and 11, along with the uncertainties
$\sigma_{\mbox{\scriptsize
ref}}$ and $\sigma_{\mbox{\scriptsize diff}}$ in columns 8 and 10. 
Instrumental magnitudes are related to the other quantities via 
$m_{\mbox{\scriptsize ins}} = 25.0 - 2.5 \log ( f_{\mbox{\scriptsize ref}} +
f_{\mbox{\scriptsize diff}}/p)$.
This is an extract
from the full table, which is available only with the electronic version of the
article.}
\centering
\begin{tabular}{ccccccccccc}
\hline
Variable &Filter & HJD & $M_{\mbox{\scriptsize std}}$ &
$m_{\mbox{\scriptsize ins}}$
& $\sigma_{m}$ & $f_{\mbox{\scriptsize ref}}$ & $\sigma_{\mbox{\scriptsize ref}}$ &
$f_{\mbox{\scriptsize diff}}$ &
$\sigma_{\mbox{\scriptsize diff}}$ & $p$ \\
Star ID  &        & (d) & (mag)                        & (mag)                       
& (mag)        & (ADU s$^{-1}$)               & (ADU s$^{-1}$)                    &
(ADU s$^{-1}$)                &
(ADU s$^{-1}$)                     &     \\
\hline
V1& V & 2455987.37467&15.391&16.588& 0.003& 2628.185&11.113& $-$317.143&7.107&1.0195\\
V1& V & 2455987.37903&15.379&16.575& 0.003& 2628.185&11.113& $-$275.868&7.261&0.9697\\
\vdots   & \vdots & \vdots  & \vdots & \vdots & \vdots & \vdots   & \vdots & \vdots &
\vdots & \vdots \\
V1& I & 2455987.37248&14.752&16.042& 0.005& 4039.892&25.259&$-$222.036&19.331&1.0588\\
V1& I & 2455987.37686&14.739&16.029& 0.006& 4039.892&25.259&$-$171.172&22.740&1.0423\\
\vdots   & \vdots & \vdots  & \vdots & \vdots & \vdots & \vdots   & \vdots & \vdots  &
\vdots & \vdots \\
V3& V & 2455987.37467&14.938&16.128& 0.002& 2771.760&11.054& +780.793&7.651&1.0195\\
V3& V & 2455987.37903&14.949&16.139& 0.002& 2771.760&11.054& +710.258&7.721&0.9697\\
\vdots   & \vdots & \vdots  & \vdots & \vdots & \vdots & \vdots   & \vdots & \vdots  &
\vdots & \vdots \\
V3& I & 2455987.37248&14.382&15.667& 0.004& 4583.013&25.240& +875.643&20.340&1.0588\\
V3& I & 2455987.37686&14.403&15.689& 0.005& 4583.013&25.240& +750.576&23.074&1.0423\\
\vdots   & \vdots & \vdots  & \vdots & \vdots & \vdots & \vdots   & \vdots & \vdots  &
\vdots & \vdots \\
\hline
\end{tabular}
\label{tab:vi_phot}
\end{table*}

\begin{figure*}
\includegraphics[width=13.5cm,height=17.5cm]{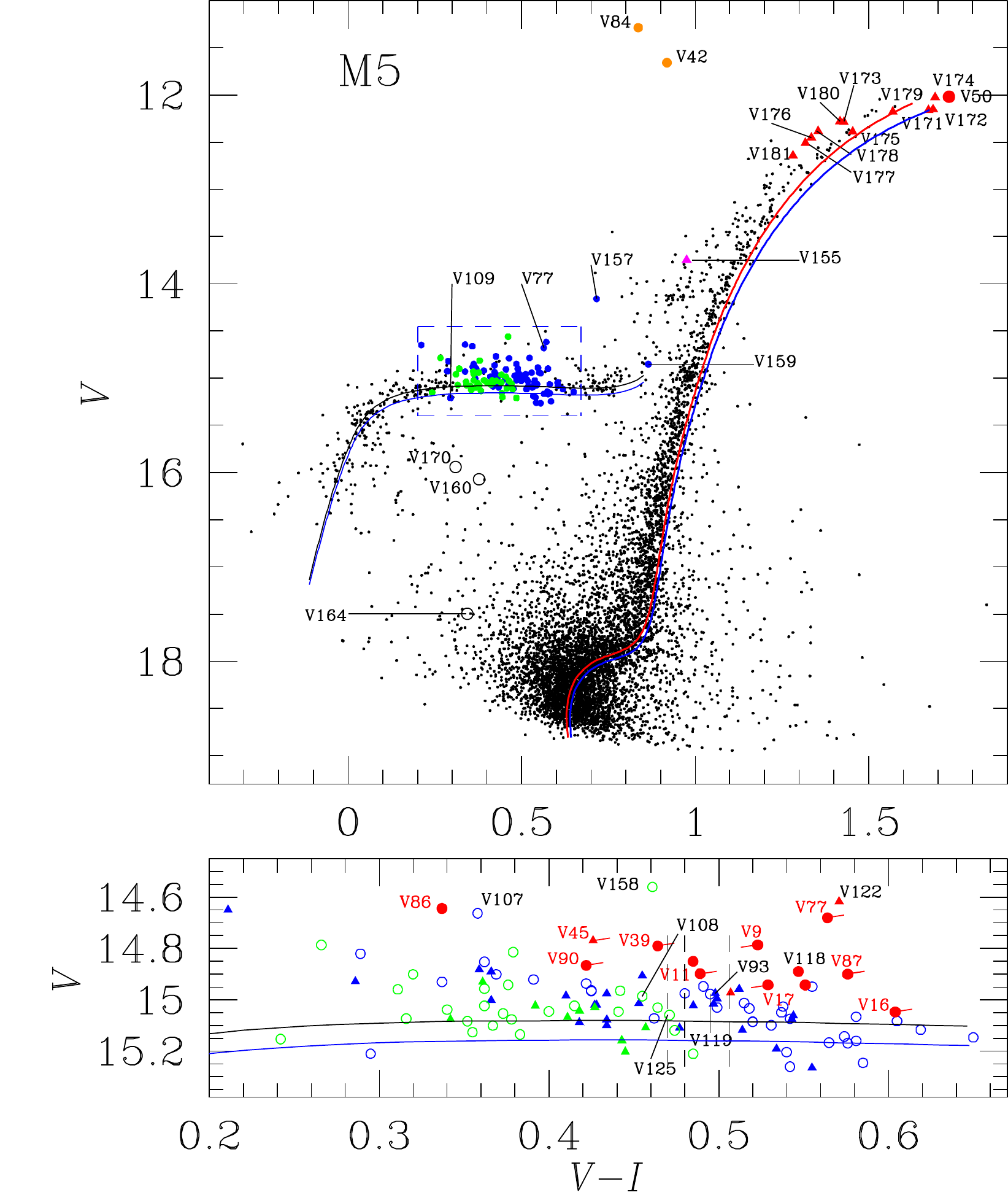}
\caption{CMD of M5. In the top panel blue and green circles represent RRab and RRc
stars respectively. Different colours code for other types of variables as follows:
semi-regular late-type (SRA) stars
red triangles; W Virginis (CW) yellow circles; SX Phe empty
black circles. Some peculiar stars are labelled and discussed in $\S$
\ref{sec:IND_STARS}. The isochrones of 12.0 Gyr for [Fe/H]=$-1.31$ (blue) and $-1.42$
(red) and the ZAHB models for [Fe/H]=$-1.42$, [$\alpha$/Fe]=0.4 and Y=0.25 (blue)
and Y=0.27 (black)
are from VandenBerg et al. (2014).
The blue box containing the RR Lyrae is expanded in the lower panel, where
stable stars are plotted as open circles, blue and green circles represent RRab
and RRc stars, and Blazhko stars are plotted as filled triangles. Stars shown as red
circles are those falling on the evolved sequence on the
Period-Amplitude diagram of Fig. \ref{Bailey} which, along with their obvious
systematically higher luminosity, identifies them as stars in a more advanced stage of
evolution towards the asymptotic giant branch (AGB). Labelled stars in black digits
are those
displaying too short a period for their colour (Fig. \ref{LP_COL}) and probably have
an unseen companion or an helium enhanced atmosphere. Stars labelled in red digits
are evolved stars discussed
individually in $\S$ \ref{sec:IND_STARS}, those with a line segment to the right or
left have increasing or decreasing periods respectively (see discussion on $\S$
\ref{summary}). Three vertical
dashed lines represent the empirical RRab-RRc border found (from left to right) in NGC
6229, NGC 5024 and NGC 4590 (see discussion on $\S$ \ref{sec:HB}).
}
\label{CMD}
\end{figure*}

\subsection{Photometric Calibrations}

\subsubsection{Relative calibration}
\label{sec:rel}

All photometric data suffer from systematic errors to some level that sometimes may be
severe enough
to be mistaken for bona fide variability in light curves. However,
multiple observations of a set of objects at different epochs, such as time-series
photometry,
may be used to investigate, and possibly correct, these systematic errors (see for
example
Honeycutt 1992). This process is a relative self-calibration of the photometry, which
is being
performed as a standard post-processing step for large-scale surveys (e.g. Padmanabhan
et al. 2008;
Regnault et al. 2009).

We apply the methodology developed in Bramich \& Freudling (2012) to solve for the
magnitude offsets
$Z_{k}$ that should be applied to each photometric measurement from the image $k$. In
terms of DIA,
this translates into a correction (to first order) for the systematic error introduced
into the photometry
from an image due to an error in the fitted value of the photometric scale factor $p$
(Bramich et al. 2015).
We found that, for either filter, the magnitude offsets that we derive are of the
order of $\sim0.02$ mag and $\sim0.03$ mag in $V$ and $I$, respectively. Applying
these magnitude offsets to our DIA photometry improves the light curve quality,
especially for the brighter stars.

\subsubsection{Absolute calibration}
\label{absolute}

Standard stars in the field of M5 are included in the online collection of
Stetson (2000)\footnote{
http://www3.cadc-ccda.hia-iha.nrc-cnrc.gc.ca/community/STETSON/standards}
and we used them to transform instrumental $vi$ magnitudes into the standard
\emph{VI} system. 

The standard minus the instrumental magnitudes show mild dependencies on the colour,
as can be seen in Fig.\ref{transV}. The transformations are of the form 

\begin{equation}
V_{std}= v +0.058(\pm0.008)(v-i) - 1.219(\pm0.007),
\label{eq:transV}
\end{equation}
\begin{equation}
I_{std}= i +0.037(\pm0.008)(v-i) - 1.305(\pm0.007).
\label{eq:transI}
\end{equation}

All of our \emph{VI} photometry for the variable stars in the FoV of our
collection
of images of M5 is provided in Table \ref{tab:vi_phot}. A small portion of this
table is given in the printed version of this paper and the full table is available in
electronic form. Despite the fact that in the present paper we only deal with RR
Lyrae and SX Phe stars, in the electronic version of the table we have also included the
photometry of all variables listed in Table \ref{variables}.

Fig. \ref{rmsCMD} shows the rms magnitude deviation in our $V$ and $I$ light curves, 
after the application of the relative photometric calibration of
Section~\ref{sec:rel}, as a function of the mean magnitude. 

To help us discuss the variable star search and classifications, we have
built 
the colour-magnitude diagram (CMD) of Fig. \ref{CMD} by calculating the
inverse-variance weighted mean magnitudes
of ∼6890 stars with $V$ and $I$ magnitudes. For better precision, the periodic
variables like RRL and SX Phe are plotted using their intensity-weighted
 magnitudes $<V>$ and colours $<V>-<I>$. The colour $<V>-<I>$ was preferred over
$<V-I>$
since our $V$ and $I$ observations are not simultaneous and building up the $V-I$
colour
curve would imply an undesirable interpolation process. The expansion of the
HB in the
bottom panel is discussed later in sections \ref{sec:HB} and \ref{summary}.

\subsection{Astrometry}
\label{sec:astrometry}

A linear astrometric solution was derived for the
$V$ filter reference image by matching 950 hand-picked
stars with the UCAC4 star catalogue (Zacharias et al. 2013)
using a field overlay in the image display tool
{\tt GAIA}\footnote{http://star-www.dur.ac.uk/\~{}pdraper/gaia/gaia.html}
(Draper 2000). We achieved a radial RMS scatter in the residuals of
$\sim$0.17~arcsec. The astrometric fit was then used to calculate the
J2000.0 celestial coordinates for all of the confirmed variables in our
FoV (see Table~\ref{variables}). The coordinates correspond to the epoch
of the $V$
reference image which pertains to the average heliocentric Julian day of
the six images used to form the reference image, 2456061.37~d.

\begin{table*}
\footnotesize
\caption{General data for all of the variables in M5 in the FoV of 
our images. All variables with Blazhko modulations are labeled '$Bl$'. 
Amplitudes for Blazhko variables correspond to the maximum observed.}
\label{variables}
\centering
\begin{tabular}{llllllclcll}
\hline
Variable & Variable & $<V>$   & $<I>$   & $A_V$       & $A_I$   & $P$ (days)    &
 HJD$_{\rm max}$     &  RA          & Dec.         \\
Star ID  & Type$^{a}$& (mag)   & (mag)   & (mag)       & (mag)   & this work     
&  (d +245~0000.)    &  (J2000.0)   & (J2000.0)  \\
\hline
V1 & RRab $Bl^{b}$ & 15.165 & 14.639 & 1.06 & 0.70 & 0.521794 &   6757.2390 &
15:18:35.46 & +02:07:29.5 \\
V3 & RRab & 15.076 & 14.486 & 0.73 & 0.47 & 0.600189 &   6029.2645 &
15:18:44.22 & +02:06:38.2 \\
V4 & RRab $Bl^{b}$ & 15.170 & 14.703 & 1.19 & 0.79 & 0.449647 &   6046.2142 & 
15:18:32.59 & +02:06:03.9 \\
V5 & RRab $Bl^{b}$ & 15.150 & 14.567 & 1.11 & 0.70 & 0.545853 &   6061.3453 &
15:18:32.85 & +02:05:41.8 \\
V6 & RRab & 15.149 & 14.626 & 0.98 & 0.69 & 0.548828 &   5989.4522 &
15:18:34.99 & +02:04:02.4 \\
V7 & RRab & 15.136 & 14.613 & 1.12 & 0.74 & 0.494413 &   5989.4323 &
15:18:32.52 & +02:01:38.8 \\
V8 & RRab $Bl^{b}$ & 15.098 & 14.541 & 0.89 & 0.57 & 0.546251 &   5989.3293 &
15:18:41.95& +02:02:32.5 \\
V9 & RRab & 14.907 & 14.343 & 0.77 & 0.51 & 0.698899 &   6063.4105 &
15:18:46.50 & +02:06:11.5 \\
V11 & RRab & 14.986 & 14.468 & 1.14 & 0.74 & 0.595897 &   6046.2751 &
15:18:23.10 & +02:06:19.7 \\
V12 & RRab & 15.154 & 14.676 & 1.27 & 0.84 & 0.467699 &   6046.2434 &
15:18:21.50 & +02:04:38.9 \\
V13 & RRab $Bl$ & 15.102 & 14.576 & 1.13 & 0.78 & 0.513133 &   6046.2579 &
15:18:33.88 & +02:03:44.4 \\
V14 & RRab $Bl^{b}$ & 15.138 & 14.646 & 1.22 & 0.83 & 0.487156 &   6063.2014 &
15:18:23.74 & +02:06:38.4 \\
V15 & RRc & 15.052 & 14.628 & 0.40 & 0.27 & 0.336765 &  6061.4425 &
15:18:46.13 & +02:04:47.4 \\
V16 & RRab & 14.911 & 14.374 & 1.21 & 0.79 & 0.647632 &  6756.4764 &
15:18:39.53 & +02:06:10.6 \\
V17 & RRab & 14.970 & 14.433 & 1.16 & 0.75 & 0.601390 &  6312.5181 &
15:18:31.61 & +02:05:35.1 \\
V18 & RRab $Bl^{b}$ & 15.119 & 14.659 & 1.22 & 0.82 & 0.463961 &  6757.2450&
15:18:43.19 & +02:02:57.3 \\
V19 & RRab $Bl^{b}$ & 15.119 & 14.665 & 1.18 & 0.82 & 0.469999 &  6046.2639 &
15:18:48.80 & +02:02:33.0 \\
V20 & RRab & 15.047 & 14.464 & 0.95 & 0.59 & 0.609473 &  5987.4631 &
15:18:16.13 & +02:04:34.1 \\
V24 & RRab $Bl^{b}$ & 15.121 & 14.663 & 1.14 & 0.74 & 0.478439 &  6029.3187 &
15:18:29.98 & +02:03:40.0 \\
V25 & RRab $Bl$ & 15.169 & 14.593 & 0.79 & 0.50 & 0.507525 &  5987.3835 &
15:18:30.98 & +02:02:42.5 \\
V26 & RRab $Bl$ & 15.008 & 14.463 & 1.05 & 0.71 & 0.622561 &  5989.4356 &
15:18:34.93 & +02:06:30.5 \\
V27 & RRab $Bl^{b}$ & 15.133 & 14.612 & 0.82 & 0.53 & 0.4703217$^{c}$ &  6063.3866 &
15:18:32.68 & +02:03:51.1 \\
V28 & RRab $Bl$ & 15.114 & 14.561 & 1.12 & 0.71 & 0.543877 &  6312.5122 &
15:18:41.86 & +02:02:44.8 \\
V30 & RRab $Bl^{b}$ & 15.085 & 14.497 & 0.82 & 0.54 & 0.592178 &  5987.4673 &
15:18:34.35 & +02:01:16.3 \\
V31 & RRc & 15.078 & 14.701 & 0.50 & 0.33 & 0.300580 &  6046.2639 &
15:18:43.12 & +02:02:23.4 \\
V32 & RRab & 15.118 & 14.649 & 1.22 & 0.82 & 0.457785 &  5989.4255 &
15:18:46.46 & +02:02:13.1 \\
V33 & RRab & 15.145 & 14.636 & 1.12 & 0.72 & 0.501481 &   5989.3506 &
15:18:32.11 & +02:06:57.8 \\
V34 & RRab & 15.127 & 14.567 & 0.83 & 0.57 & 0.568142 &  6061.2497 &
15:18:39.02 & +02:05:46.6 \\
V35 & RRc $Bl$ & 14.959 & 14.607 & 0.48 & 0.29 & 0.308217 &  6063.2209 &
15:18:32.21& +02:02:55.7 \\
V36 & RRab & 15.060 & 14.484 & 0.72 & 0.44 & 0.627725 &  6757.2850 &
15:18:32.66 & +02:03:58.9 \\
V37 & RRab & 15.105 & 14.589 & 1.01 & 0.63 & 0.488801 &  6029.3187 &
15:18:36.11 & +02:03:41.5 \\
V38 & RRab $Bl^{b}$& 15.166 & 14.659 & 1.03 & 0.67 & 0.470422 &  6504.2422 &
15:18:30.55 & +02:06:48.5 \\
V39 & RRab & 14.962 & 14.427 & 1.12 & 0.75 & 0.589037 &  6063.3327 &
15:18:24.35 & +02:00:44.0 \\
V40 & RRc $Bl$ & 15.069 & 14.656 & 0.46 & 0.28 & 0.317327 &  6063.1929 &
15:18:41.85& +02:06:39.0 \\
V41 & RRab & 15.110 & 14.619 & 1.29 & 0.83 & 0.488572 &  6061.3027 &
15:18:35.04 & +02:08:39.9 \\
V42 & CW & 11.659 &10.740 & 1.32 & 0.90 &25.735  & 6046.2397  &
15:18:24.80 & +02:02:53.5 \\
V43 & RRab & 15.029 & 14.428 & 0.56 & 0.43 & 0.660226 &  6757.4831 &
15:18:20.09 & +02:07:30.9 \\
V44 & RRc $Bl$ & 15.091 & 14.633 & 0.45 & 0.26 & 0.329599 &  5987.4916 &
15:18:26.47 & +02:05:24.4 \\
V45 & RRab $Bl$ & 15.042 & 14.471 & 1.01 & 0.66 & 0.616636 &  6061.3946 &
15:18:25.59 & +02:05:59.5 \\
V47 & RRab & 15.153 & 14.595 & 0.94 & 0.61 & 0.539730 &  6757.2200 &
15:18:28.35 & +02:05:50.5 \\
V50 & SRA &12.15  &10.27  &0.73  &0.27 &107.6 & 6061.4267 &15:18:36.04& +02:06:37.8\\
V52 & RRab $Bl^{b}$& 15.011 & 14.547 & 1.07 & 0.70 & 0.501541 &   6029.2645 & 
15:18:40.56 & +02:05:21.8 \\
V53 & RRc $Bl$ & 14.861 & 14.455 & 0.46 & 0.30 & 0.373519 &   5987.5209 &
15:18:37.92 & +02:05:06.8 \\
V54 & RRab & 15.183 & 14.718 & 1.27 & 0.90 & 0.454115 &   5989.3176 & 
15:18:35.41 & +02:05:46.0 \\
V55 & RRc $Bl$ & 15.086 & 14.634 & 0.41 & 0.26 & 0.328903 &  6504.1925 &
15:18:38.29 & +02:02:04.1 \\
V56 & RRab $Bl^{b}$ & 15.127 & 14.585 & 1.01 & 0.69 & 0.534690 &  6061.3186 &
15:18:28.86 & +02:06:28.6 \\
V57 & RRc & 15.100 & 14.741 & 0.50 & 0.32 & 0.284697 &  6046.2434 &
15:18:31.43 & +02:06:30.5 \\
V59 & RRab & 15.079 & 14.537 & 0.99 & 0.65 & 0.542025 &  6061.2807 &
15:18:23.15 & +02:04:19.8 \\
V60 & RRc & 15.115 & 14.732 & 0.51 & 0.33 & 0.285236 &  6504.1518 &
15:18:25.94 & +02:05:01.9 \\
V61 & RRab & 15.113 & 14.544 & 0.92 & 0.61 & 0.568642 &  6061.4250 &
15:18:16.15 & +02:04:27.7 \\
V62 & RRc & 15.078 & 14.733 & 0.49 & 0.33 & 0.281417 &  5989.3176 & 
15:18:43.98 & +02:01:07.9 \\
V63 & RRab $Bl^{b}$ & 15.110 & 14.627 & 1.10 & 0.66 & 0.497686 &  6756.3534 &
15:18:47.61 & +02:05:34.9 \\
V64 & RRab & 15.114 & 14.559 & 0.96 & 0.63 & 0.544489 &  6062.1833&
15:18:29.32 & +02:00:42.6 \\
V65 & RRab $Bl^{b}$ & 15.117 & 14.625 & 1.14 & 0.74 & 0.480664 &  5989.4389 &
15:18:22.37 & +02:03:21.8 \\
V74 & RRab & 15.155 & 14.674 & 1.33 & 0.97 & 0.453984 &  6061.3915 &
15:18:47.19 & +02:07:25.7 \\
V77 & RRab & 14.744 & 14.148 & 0.57 & 0.44 & 0.845158 &  6061.3518 &
15:18:21.40 & +02:01:50.9 \\
V78 & RRc & 15.117 & 14.778 & 0.39 & 0.25 & 0.264820 &  5989.5204 &
15:18:37.97 & +02:07:27.0 \\
V79 & RRc & 15.018 & 14.564 & 0.38 & 0.24 & 0.333139 &  6046.2326 &
15:18:24.25 & +02:04:22.5 \\
V80 & RRc & 15.095 & 14.629 & 0.39 & 0.25 & 0.336542 &  6046.2751 &
15:18:30.24& +02:06:42.9 \\
\hline
\hline
\end{tabular}
\end{table*}

\begin{table*}
\footnotesize
\addtocounter{table}{-1}
\caption{Continued}
\label{variablesB}
\centering
\begin{tabular}{llllllclcll}
\hline
Variable & Variable & $<V>$   & $<I>$   & $A_V$       & $A_I$   & $P$ (days)    &
 HJD$_{max}$     &  RA          & Dec.         \\
Star ID  & Type$^{a}$ & (mag)   & (mag)   & (mag)       & (mag)   & this work & (d
 +245~0000.)    & (J2000.0)   & (J2000.0)    \\
\hline
V81 & RRab & 15.098 & 14.538 & 0.95 & 0.60 & 0.557271 &  6504.2422 &
15:18:28.18 & +02:02:50.6\\
V82 & RRab & 15.084 & 14.512 & 0.90 & 0.60 & 0.558435 &  6063.2014 &
15:18:28.76 & +02:05:04.7 \\
V83 & RRab & 15.122 & 14.572 & 0.86 & 0.59 & 0.553307 &  6061.4320 &
15:18:27.41 & +02:03:25.0\\
V84 & CW & 11.287 & 10.451 & 0.97 &0.84 & 26.49 &  6754.0000 &
15:18:36.13 & +02:04:16.7\\
V85 & RRab $Bl$ & 14.996 & 14.523 & 0.85 & 0.57 & 0.527535 &  6061.3804&
15:18:35.75 & +02:04:14.3 \\
V86 & RRab & 14.944 & 14.439 & 1.24 & 0.87 & 0.567513 &  6504.1925 &
15:18:35.50 & +02:04:15.9 \\
V87 & RRab & 14.954 & 14.349 & 0.35 & 0.25 & 0.738421 &  6061.2186 &
15:18:41.42 & +02:04:44.1 \\
V88 & RRc & 15.056 & 14.651 & 0.42 & 0.27 & 0.328090 &  5989.5102 &
15:18:37.75 & +02:05:49.5 \\
V89 & RRab & 15.126 & 14.570 & 0.94 & 0.63 & 0.558443 & 6063.1970 &
15:18:37.41 & +02:05:52.5 \\
V90 & RRab & 15.027 & 14.496 & 1.30 & 0.84 & 0.557168 &  6061.3518 &
15:18:30.31 & +02:05:06.7\\
V91 & RRab & 15.097 & 14.518 & 0.82 & 0.55 & 0.584945 &  6063.4183&
15:18:30.93 & +02:05:26.3 \\
V92 & RRab $Bl$ & 15.146 & 14.637 & 1.25 & 0.86 & 0.463388 &  6061.1878 &
15:18:29.22 & +02:02:48.4 \\
V93 & RRab & 15.229 & 14.549 & 1.33 & 0.85 & 0.552300 &  5987.4505 &
15:18:36.12 & +02:04:13.0 \\
V94 & RRab & 15.193 & 14.628 & 1.05 & 0.69 & 0.531327 &  6061.2497 &
15:18:31.71 & +02:05:08.1 \\
V95 & RRc & 15.050 & 14.675 & 0.50 & 0.35 & 0.290832 &  6061.3613 &
15:18:30.33 & +02:06:34.0 \\
V96 & RRab & 15.157 & 14.640 & 1.0 & 0.7 & 0.512255 &  6312.48 &
15:18:32.50 & +02:05:23.3 \\
V97 & RRab $Bl^{b}$ & 15.115 & 14.566 & 0.90 & 0.60 & 0.544656 &  5987.3747 &
15:18:36.33 & +02:03:16.0 \\
V98 & RRc & 15.094 & 14.674 & 0.47 & 0.30 & 0.306360 &  6063.4216 &
15:18:35.81 & +02:05:08.6 \\
V99 & RRc $Bl$ & 15.093 & 14.673 & 0.52 & 0.34 & 0.321336 &  6061.2186 &
15:18:35.57 & +02:04:48.7 \\
V100 & RRc & 15.146 & 14.769 & 0.50 & 0.33 & 0.294365 &  6504.1849 &
15:18:33.55 & +02:05:38.6 \\
V101 & U~Gem & $>19.0$ &$>19.0$  &--& -- &--&--&15:18:14.51& +02:05:35.7 \\
V102 & RRab $Bl$ & 14.621 & 14.410 & 1.16 & 0.84 & 0.470540 &   6061.3152 &
15:18:34.37 & +02:04:34.4 \\
V103 & RRab & 15.074 & 14.523 & 0.83 & 0.56 & 0.566660 &   5989.4489 &
15:18:34.63 & +02:04:40.7 \\
V104 & RRab & 15.072 & 14.614 & 0.80 & 0.49 & 0.486748 &   6504.1830 &
15:18:32.68 & +02:05:32.4 \\
V105 & RRc & 15.234 & 14.957 & 0.58 & 0.48 & 0.295025 &   5987.3961 & 
15:18:32.88 & +02:05:05.5 \\
V106 & RRab $Bl$ & 15.192 & 14.703 & 1.25 & 0.96 & 0.527383 &  5987.5136 &
15:18:32.93 & +02:04:59.5 \\
V107 & RRab & 14.932 & 14.419 & 1.03 & 0.69 & 0.511698 &  5987.4799 &
15:18:33.25 & +02:04:46.0 \\
V108 & RRc & 14.980 & 14.517 & 0.48 & 0.31 & 0.328628 &  5989.5204 &
15:18:33.79 & +02:04:47.0 \\
V109 & RRab & 15.549 & 14.992 & 2.0 & 1.4 & 0.473008 &  6046.3241 &
15:18:34.57 & +02:04:50.8 \\
V110 & RRab & 15.254 & 14.706 & 0.72 & 0.54 & 0.597996 &  6312.5083 &
15:18:34.88 & +02:04:45.7\\
V111 & RRab & 15.074 & 14.470 & 0.85 & 0.50 & 0.634647 &  5989.4221 &
15:18:34.89 & +02:05:05.5 \\
V112 & RRab $Bl$ & 15.074 & 14.592 & 0.8 & 0.6 & 0.534456 &  5989.3749 &
15:18:35.11 & +02:04:17.4 \\
V113 & RRc & 15.101 & 14.726 & 0.5 & 0.3 & 0.284676 &  5989.3541 &
15:18:35.12 & +02:04:15.2 \\
V114 & RRab & 15.167 & 14.592 & 0.84 & 0.58 & 0.603659 &  6504.1668 &
15:18:35.24 & +02:04:46.5 \\
V115 & RRab & 15.045 & 14.445 & 0.59 & 0.41 & 0.609084 &  5989.3644 &
15:18:36.33& +02:04:53.8 \\
V116 & RRc & 14.972 & 14.579 & 0.47 & 0.31 & 0.347289 &  5987.5209 &
15:18:36.33 & +02:04:46.1 \\
V117 & RRc & 14.951 & 14.512 & 0.35 & 0.24 & 0.335929 &  6061.4320 &
15:18:34.90 & +02:04:52.2 \\
V118 & RRab & 14.870 & 14.294 & 1.16 & 0.71 & 0.580517 &  6063.2638 &
15:18:34.59 & +02:04:57.0 \\
V119 & RRab & 15.171 & 14.531 & 0.94 & 0.58 & 0.550962 &  6504.2405 &
15:18:34.13 & +02:05:05.1 \\
V120 & RRc $Bl$ & 15.154 & 14.737 & 0.58 & 0.39 & 0.278719 &  5989.5137 &
15:18:33.83 & +02:04:57.6\\
V121 & RRab $Bl$ & 15.348 & 14.623 & 1.15 & 0.66 & 0.599039 &  6063.2528 &
15:18:32.93 & +02:04:43.0 \\
V122 & RRab $Bl$ & 14.682 & 14.077 & 0.64 & 0.51 & 0.733089 &  5987.4589 & 
15:18:32.02 & +02:04:45.1\\
V123 & RRab & 15.167 & 14.530 & 0.68 & 0.45 & 0.602486 &  6504.1686 &
15:18:33.26 & +02:05:18.8 \\
V125 & RRc & 15.098 & 14.594 & 0.44 & 0.28 & 0.303304 &  6063.2865 &
15:18:32.17 & +02:05:06.8 \\
V126 & RRc & 15.147 & 14.659 & 0.45 & 0.40 & 0.343258 &  5989.3506 &
15:18:32.06& +02:05:02.2 \\
V127 & RRab $Bl$ & 14.673 & 14.548 & 0.81 & 0.77 & 0.540366 & 
5987.4839 & 15:18:30.76 & +02:05:00.8 \\
V128 & RRc & 15.109 & 14.709 & 0.48 & 0.32 & 0.306013 &  5989.4122 &
15:18:30.81& +02:04:42.5\\
V129 & RRab & 15.136 & 14.574 & 0.59 & 0.41 & 0.605302 &  6504.2332 &
15:18:30.05 & +02:04:01.7 \\
V130 & RRc $Bl$ & 14.993 & 14.605 & 0.6 & 0.4 & 0.327187 &  6046.2288 &
15:18:38.59 & +02:05:44.7\\
V131 & RRc $Bl$ & 15.149 & 14.774 & 0.6 & 0.4 & 0.281533 &  6046.2326 &
15:18:38.60& +02:05:42.2\\
V132 & RRc & 15.035 & 14.678 & 0.38 & 0.26 & 0.283738 &  6029.4767&
15:18:36.28 & +02:05:33.8 \\
V133 & RRc & 14.989 & 14.606 & 0.47 & 0.28 & 0.294864 &  5987.3747 &
15:18:40.53 & +02:05:30.1 \\
V137 & RRab & 15.159 & 14.558 & 0.50 & 0.36 & 0.619359 &  6063.2901 &
15:18:36.33 & +02:05:28.8 \\
V139 & RRc & 14.796 & 14.517 & 0.37 & 0.26 & 0.300356 &  6504.1518 &
15:18:32.09 & +02:05:22.8 \\
V142 & RRab $Bl$ & 15.087 & 14.714 & 1.4 & 1.0 & 0.458151 &  6029.3187 &
15:18:34.48 & +02:04:50.2 \\
V155 & EW &13.74  &12.76 &0.14  &0.09 &0.664865&6504.2067  &15:18:33.40
&+02:05:12.2\\
\hline
\hline
\end{tabular}
\end{table*}

\begin{table*}
\footnotesize
\addtocounter{table}{-1}
\caption{Continued}
\label{variablesB}
\centering
\begin{tabular}{llllllclcll}
\hline
Variable & Variable & $<V>$   & $<I>$   & $A_V$       & $A_I$   & $P$ (days)    &
 HJD$_{max}$     &  RA          & Dec.         \\
Star ID  & Type$^{a}$& (mag)   & (mag)   & (mag)       & (mag)   & this work    
& (d
 +245~0000.)    & (J2000.0)   & (J2000.0)    \\
\hline
V156 & RRab $^{d}$& 12.912 & 11.721 & 0.18 & -- & -- &  -- &
15:18:32.63 & +02:04:59.0 \\
V157 & RRab & 14.140 & 13.434 & 0.60 & -- & 0.517608 &  6063.2014 &
15:18:33.37 & +02:04:58.0 \\
V158 & RRc & 14.580 & 14.106 & 0.45 & 0.31 & 0.442627 &  5989.4752 &
15:18:32.83 & +02:04:50.6\\
V159 & E &14.85  &13.99& -- &-- &-- &  -- &15:18:32.88& +02:04:36.5 \\
V160 & SX Phe & 16.05 & 15.70 & 0.55 & 0.34 & 0.089749 & 5989.5102&
15:18:29.84 & +02:04:09.8  \\
V161 & RRc & 15.161 & 14.702 & 0.42 & 0.26 & 0.331266 &  6046.2468 &
15:18:33.71 & +02:05:41.5 \\
V164 & SX Phe & 17.50 & 17.15 & 0.15 & -- & 0.042134 &  6029.4832&
15:18:22.77 & +02:02:49.3  \\
V170 & SX Phe & 15.95 & 15.63 & 0.57 & 0.41 & 0.089467 &  6063.3361 &15:18:32.14
& +02:04:20.4  \\
V171 & SRA & 12.17 & 10.50 & 0.25 & 0.14 & 28.8 &  6312.5083 &15:18:34.26
& +02:04:24.2  \\
V172 & SRA & 12.15 & 10.47 & 0.23 & 0.13 & -- & -- & 15:18:31.59 & +02:04:41.4
 \\
V173 & SRA & 12.28 & 10.86 & 0.13 & 0.13 & 43.1 &  6504.1686 & 15:18:28.42 &
+02:04:29.8  \\
V174 & SRA & 12.03 & 10.33 & 0.33 & 0.15 & 80.6 &  6063.4183 & 15:18:34.18 &
+02:06:25.5  \\
V175 & SRA & 12.40 & 10.94 & 0.18 & 0.13 & -- & -- &  15:18:36.22 & +02:05:11.3 
\\
V176 & SRA & 12.46 & 11.13 & 0.22 & 0.20 & 133.3 &  5989.3064 & 15:18:37.38 &
+02:06:08.2  \\
V177 & SRA & 12.51 & 11.19 & 0.13 & 0.10 & -- &  -- &  15:18:41.40 & +02:06:00.9\\
V178 & SRA & 12.39 & 11.03 & 0.12 & 0.10 & 141.6 &  5987.4759 &  15:18:33.10 &
+02:04:58.0  \\
V179 & SRA & 12.18 & 10.61 & 0.12 & 0.11 & -- &  -- & 15:18:33.42 &
+02:04:59.6  \\
V180 & SRA & 12.27 & 10.86 & 0.24 & 0.24 & -- &-- & 15:18:35.82 &
+02:03:42.4  \\
V181 & SRA & 12.64 & 11.36 & 0.07 & 0.08 & -- & --&  15:18:45.40 &
+02:04:30.9  \\
\hline
\hline
\end{tabular}
\tablenotetext {a}{All variable types follow the designations from the
General Catalogue of Variable Stars (Samus et al. 2009). Or see also the AAVSO link
http://www.aavso.org/vsx/help/VariableStarTypeDesignationsInVSX.pdf}
\tablenotetext {b}{Blazhko variations identified by Jurcsik et al. (2011).}
\tablenotetext{c}{Adopted from Szeidl et al. (2011).} 
\tablenotetext {d}{Classification based on the light curve of Drissen \& Shara
(1998).}
\end{table*}

\begin{figure*} 
\includegraphics[width=15.0cm,height=22.cm]{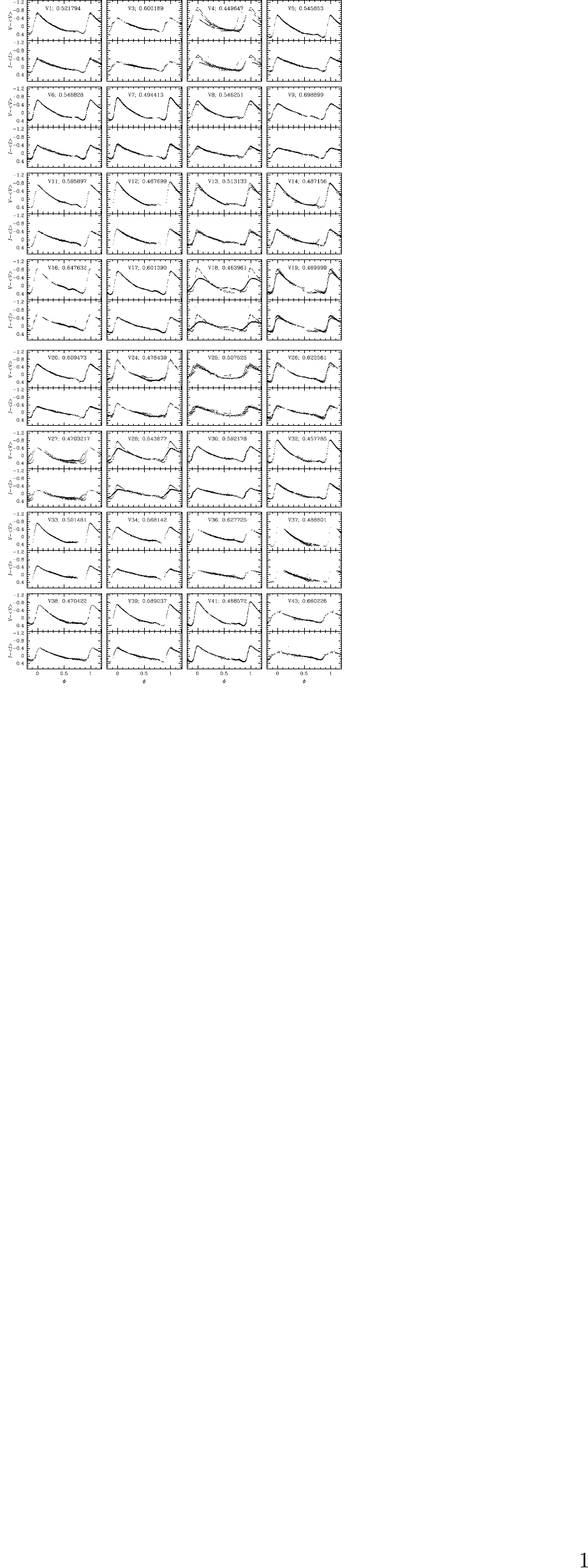}
\caption{Light curves of the RRab stars in M5. 
To preserve the vertical scale and limits, the vertical axes display 
$V-<V>$ and $I-<I>$. All light curves are
phased with the ephemerides listed in Table~\ref{variables}. Individual mean values and
amplitudes are also listed in the table.}
   \label{VARSabA}
\end{figure*}

\begin{figure*} 
\addtocounter{figure}{-1}
\includegraphics[width=15.cm,height=22.cm]{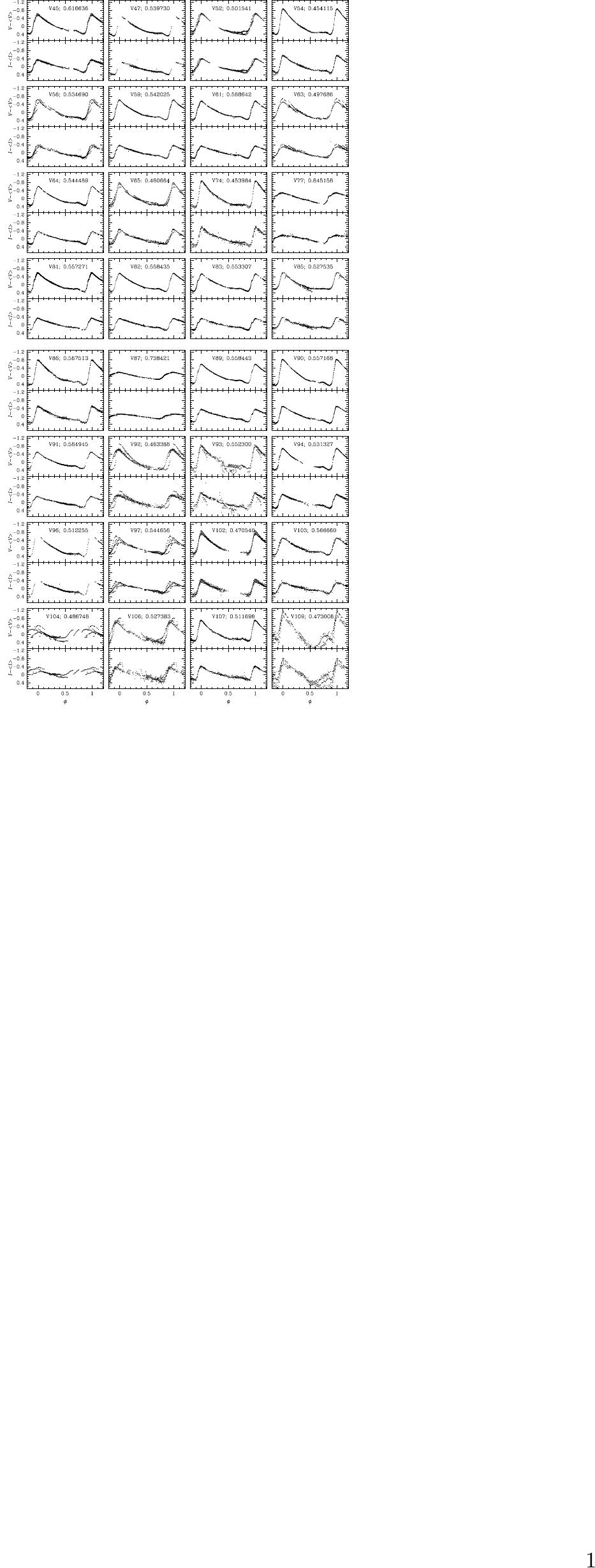}
\caption{Continued}
\end{figure*}

\begin{figure*} 
\addtocounter{figure}{-1}
\includegraphics[width=15.cm,height=11.0cm]{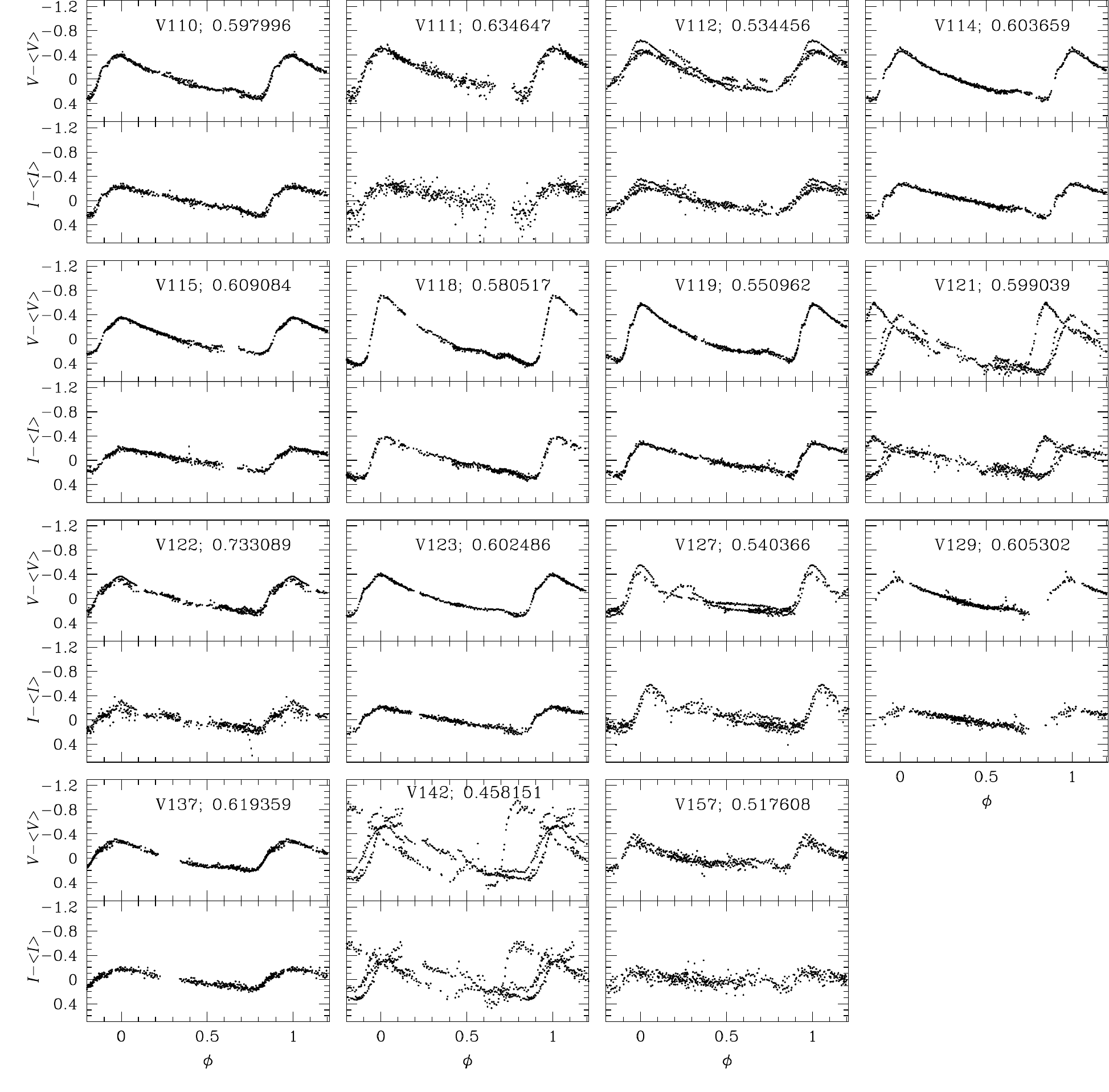}
\caption{Continued}
\end{figure*}

\begin{figure*} 
\includegraphics[width=15.cm,height=22.cm]{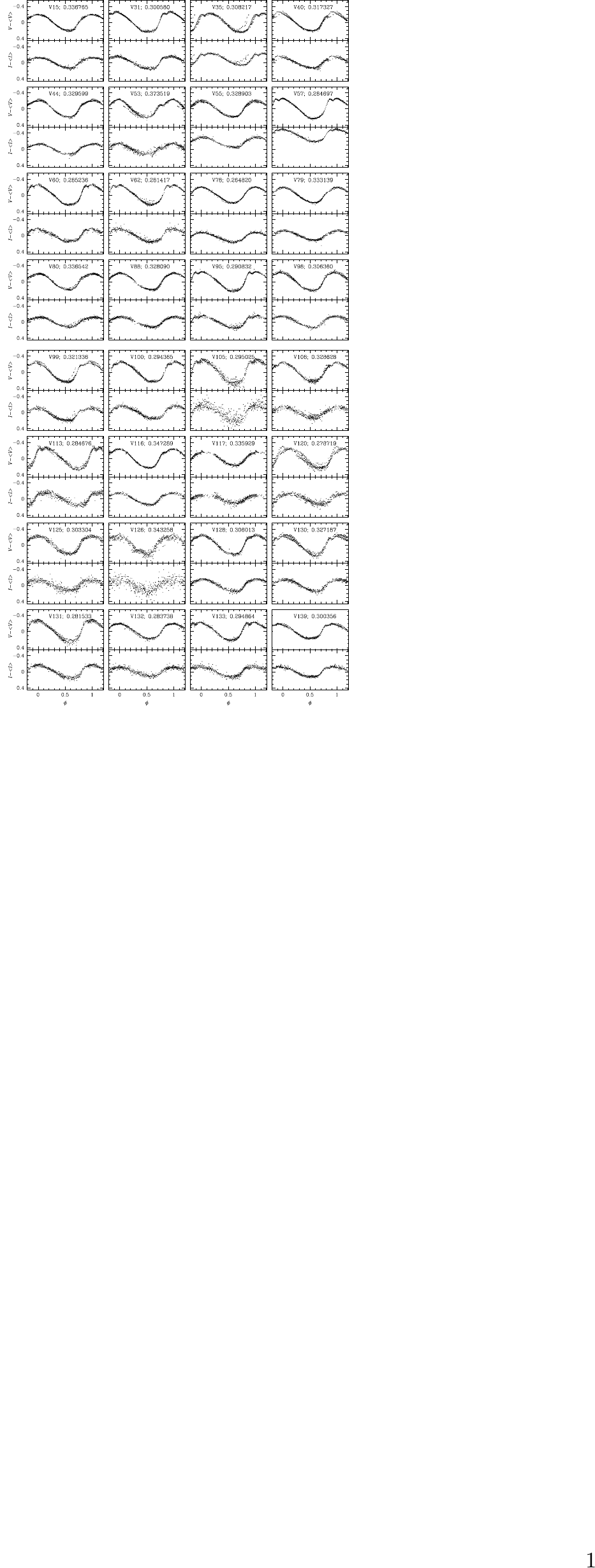}
\caption{Light curves of the RRc stars in M5
phased with the ephemerides listed in Table~\ref{variables}.}
    \label{VARSc}
\end{figure*}

\begin{figure*} 
\addtocounter{figure}{-1}
\includegraphics[width=8.cm,height=10.5cm]{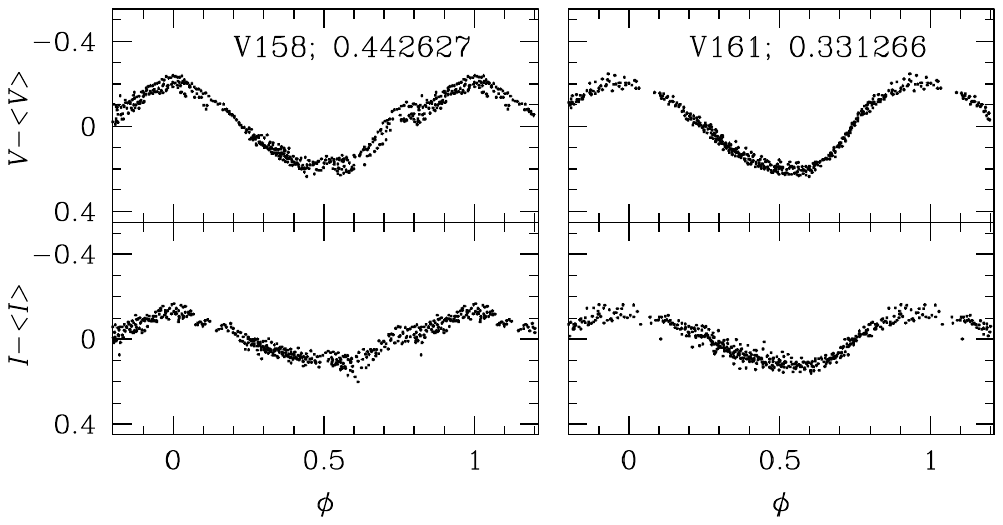}
\caption{Continued}
\end{figure*}

\section{The RR Lyrae stars in M5}
The $V$ and $I$ light curves of 79 RRab and 34 RRc stars in the FoV of our 
images are displayed in Figs. \ref{VARSabA} and \ref{VARSc}. 
Note that the variables V2, V10, V21, V22, V29, V58, V66-V73, V75, V76, V141 and 
V165-V169 are outside the FoV of our study. We do not plot the RRab star V156 in
Figure 4 because we were unable to phase our light curve, although we confirm that it
is variable (see the comment in $\S$ \ref{sec:IND_STARS} and the light curve
in Fig.\ref{V156}). The stars in Figs. \ref{VARSabA} and \ref{VARSc} are
discussed individually in $\S$ \ref{sec:IND_STARS} when found peculiar. The RR Lyrae
stars were
selected for the Fourier decomposition approach for the determination of their
physical parameters only when their light curves are considered sufficiently stable,
i.e. with no clear signs of amplitude modulations.

\subsection{Periods}
\label{sec:Periods}

The light curves of the RRab and RRc stars in Figs. \ref{VARSabA} and
\ref{VARSc} were phased with the ephemerides in Table \ref{variables}. 

The periods reported in Table \ref{variables} were calculated exclusively on our 
data from 2012-2014, applying the
string-length method (Burke, Rolland \& Boy 1970; Dworetsky 1983) as well as
{\tt period04} (Lenz \& Breger 2005). It is well known, however, from the pioneering
work
of 
Oosterhoff (1941) to the most recent analysis by Szeidl et al. (2011),
that most RRL in M5 do exhibit secular period variations, thus the periods
given in Table \ref{variables} are not likely to be the most precise ones.
Nevertheless,
they are instantaneous accurate values that phase the light curves during 2012-2014
very well, as can be judged from the phased light curves in Figs. \ref{VARSabA} and
\ref{VARSc}, and that can be accurately used to Fourier decompose the light curves in
pursuit of physical parameters, as will be described later in this paper.
In a separate paper we shall revisit the secular period change nature of the RRL
in M5 in the light of our observations which extend the time-base by at least
15 years.

\subsection{[Fe/H] and M$_V$ from light curve Fourier decomposition}
\label{sec:RRLstars}
 
Stellar physical parameters, such as [Fe/H], M$_V$, $T_{\rm eff}$, 
mass and radius for RRL can be calculated via the
Fourier decomposition of their $V$ light curves into its harmonics as:

\begin{equation}
m(t) = A_0 ~+~ \sum_{k=1}^{N}{A_k ~\cos~( {2\pi \over P}~k~(t-E) ~+~ \phi_k ) },
\label{eq_foufit}
\end{equation}

\noindent
where $m(t)$ is the magnitude at time $t$, $P$ is the period and $E$ the epoch. A
linear
minimization routine is used to derive the amplitudes $A_k$ and phases $\phi_k$ of
each harmonic, from which the Fourier
parameters $\phi_{ij} = j\phi_{i} - i\phi_{j}$ and $R_{ij} = A_{i}/A_{j}$ are
calculated. The mean
magnitudes $A_0$, and the Fourier light curve fitting parameters of
individual RRab and RRc stars are listed in Table
~\ref{fourier_coeffs}. In this table we have excluded stars with evident
amplitude-phase modulations, excessive noise, apparent blending or incomplete light
curves that badly disturbed the Fourier fit. The latter case is particularly true in
those light curves where the maximum or minimum are missing.

\begin{table*}
\footnotesize
\caption{Fourier coefficients $A_{k}$ for $k=0,1,2,3,4$, and phases $\phi_{21}$,
$\phi_{31}$ and $\phi_{41}$, for RRab and RRc stars. The numbers in parentheses
indicate
the uncertainty on the last decimal place. Also listed is the deviation 
parameter $D_{\mbox{\scriptsize m}}$ (see Section~\ref{sec:RRLstars}).}
\centering                   
\begin{tabular}{lllllllllr}
\hline
Variable ID     & $A_{0}$    & $A_{1}$   & $A_{2}$   & $A_{3}$   & $A_{4}$   &
$\phi_{21}$ & $\phi_{31}$ & $\phi_{41}$ 
&  $D_{\mbox{\scriptsize m}}$ \\
     & ($V$ mag)  & ($V$ mag)  &  ($V$ mag) & ($V$ mag)& ($V$ mag) & & & & \\
\hline
       &       &   &   &   & RRab stars    & &        
   &             &       \\
\hline
V3 & 15.076(1)  & 0.245(2) & 0.120(1) & 0.082(2) & 0.041(1) & 4.063(18)& 8.486(25) &
6.758(42) & 1.1 \\
V5 & 15.150(1) & 0.367(1) & 0.182(1) & 0.129(1) & 0.087(1) & 3.942(8) & 8.231(12) &
6.223(17) & 2.2 \\
V6 & 15.149(1) & 0.338(2) & 0.165(2) & 0.117(2) & 0.079(1) & 3.911(13) & 8.215(18) &
6.204(28) & 1.5 \\
V7 & 15.136(2) & 0.380(2) & 0.180(2) & 0.139(2) & 0.095(2) & 3.797(17) & 7.916(23) &
5.836(33) & 0.5 \\
V9 & 14.907(1) & 0.279(1) & 0.133(1) & 0.089(1) & 0.038(1) & 4.234(11) & 8.755(17) &
7.068(33) & 3.1 \\
V11 & 14.986(1) & 0.373(1) & 0.204(1) & 0.130(1) & 0.091(1) & 4.043(9) & 8.356(14) &
6.507(18) & 2.5 \\
V12 & 15.154(2) & 0.429(2) & 0.199(2) & 0.153(2) & 0.104(2) & 3.824(15) & 7.915(20) &
5.827(27)  & 1.3 \\
V16 & 14.911(2) & 0.401(3) & 0.218(3) & 0.134(3) & 0.092(2) & 4.130(15) & 8.418(22) &
6.751(34) & 4.8 \\
V17 & 14.970(1) & 0.387(1) & 0.207(1) & 0.129(1) & 0.094(1) & 3.998(7) & 8.274(11) &
6.413(15) & 2.3 \\
V20 & 15.047(1) & 0.306(1) & 0.164(1) & 0.108(2) & 0.067(1) & 4.139(12) & 8.584(18) &
6.851(28) & 2.1 \\
V30 & 15.085(1) & 0.268(2) & 0.140(2) & 0.094(2) & 0.053(2) & 3.985(16) & 8.377(24) &
6.551(38) & 1.0 \\
V32 & 15.118(1) & 0.419(2) & 0.198(2) & 0.146(2) & 0.093(2) & 3.850(13) & 7.862(18) &
5.804(26) & 1.6 \\
V33 & 15.145(1) & 0.382(2) & 0.175(2) & 0.136(2) & 0.090(2) & 3.794(15) & 7.905(19) &
5.763(28) & 4.1 \\
V34 & 15.127(1) & 0.288(1) & 0.138(1) & 0.093(2) & 0.055(1) & 3.988(15) & 8.295(20) &
6.413(31) & 1.1 \\
V38 & 15.166(1) & 0.382(2) & 0.174(2) & 0.114(2) & 0.071(2) & 3.843(12) & 7.982(19) &
5.980(28) & 1.3 \\
V39 & 14.962(1) & 0.374(1) & 0.198(1) & 0.127(1) & 0.090(1) & 3.984(8) & 8.264(12) &
6.316(17) & 1.6 \\
V41 & 15.110(1) & 0.444(1) & 0.209(1) & 0.157(1) & 0.101(2) & 3.803(9) & 7.973(13) &
5.858(18) & 1.2 \\
V54 & 15.183(1) & 0.451(2) & 0.206(2) & 0.154(2) & 0.098(2) & 3.786(10) & 7.886(15) &
5.832(21) & 1.8 \\
V59 & 15.079(1) & 0.327(1) & 0.169(1) & 0.114(1) & 0.079(1) & 3.949(9) & 8.265(14) &
6.279(19) & 0.9 \\
V61 & 15.113(1) & 0.312(1) & 0.156(1) & 0.105(1) & 0.071(1) & 4.005(11) & 8.335(17) &
6.433(24) & 1.8 \\
V64 & 15.114(1) & 0.320(1) & 0.157(1) & 0.112(1) & 0.076(1) & 3.894(12) & 8.167(17) &
6.152(24) & 0.9 \\
V74 & 15.155(1) & 0.447(2) & 0.208(2) & 0.159(2) & 0.106(2) & 3.883(15) & 8.007(21) &
5.916(29) & 1.2 \\
V77 & 14.744(1) & 0.226(1) & 0.089(2) & 0.036(2) & 0.022(2) & 4.528(23) & 9.270(50) &
7.521(74) & 4.8 \\
V81 & 15.098(1) & 0.314(1) & 0.157(1) & 0.111(1) & 0.071(1) & 3.936(10) & 8.260(14) &
6.299(21) & 1.6 \\
V82 & 15.084(1) & 0.307(1) & 0.152(1) & 0.107(1) & 0.068(1) & 3.941(9) & 8.262(13) &
6.321(19) & 1.8 \\
V83 & 15.122(1) & 0.291(2) & 0.146(2) & 0.104(2) & 0.064(2) & 3.956(15) & 8.291(22) &
6.343(32) & 0.9 \\
V86 & 14.944(1) & 0.419(2) & 0.222(2) & 0.137(2) & 0.104(2) & 3.872(11) & 8.191(17) &
6.145(23) & 2.7 \\
V89 & 15.126(1) & 0.321(1) & 0.162(1) & 0.112(1) & 0.072(1) & 3.973(8) & 8.291(12) &
6.355(17) & 1.6 \\
V90 & 15.027(1) & 0.448(1) & 0.227(2) & 0.149(2) & 0.102(2) & 3.904(9) & 8.232(14) &
6.157(18) & 1.9 \\
V91 & 15.097(1) & 0.281(1) & 0.144(1) & 0.098(1) & 0.057(1) & 4.044(11) & 8.409(16) &
6.507(24) & 0.8 \\
V94 & 15.193(2) & 0.360(3) & 0.178(3) & 0.131(4) & 0.084(3) & 3.842(25) & 8.164(33) &
6.055(48) & 2.2 \\
V103 & 15.074(2) & 0.292(2) & 0.146(2) & 0.106(2) & 0.052(2) & 3.909(20) & 8.285(29) &
6.421(49) & 2.2 \\
V107 & 14.932(1) & 0.345(2) & 0.168(2) & 0.132(2) & 0.081(2) & 3.784(15) & 7.905(21) &
5.710(20) & 1.7 \\
V110 & 15.254(1) & 0.259(2) & 0.117(2) & 0.084(2) & 0.040(2) & 4.222(20) & 8.744(29) &
7.024(52) & 3.9 \\
V114 & 15.167(1) & 0.285(1) & 0.134(1) & 0.094(1) & 0.052(1) & 4.056(15) & 8.336(22) &
6.549(34) & 1.8 \\
V118 & 14.870(2) & 0.386(2) & 0.208(2) & 0.134(2) & 0.093(2) & 4.018(14) & 8.354(21) &
6.234(32) & 3.7 \\
V119 & 15.171(1) & 0.311(2) & 0.156(2) & 0.110(2) & 0.073(2) & 3.929(15) & 8.200(22) &
6.286(32) & 0.8 \\
V123 & 15.167(1) & 0.244(1) & 0.117(1) & 0.074(1) & 0.035(1) & 4.118(15) & 8.528(23) &
6.817(42) & 0.7 \\
\hline
             &            &           &           &           & RRc stars &           
 &             &             &\\
\hline
V15 & 15.052(1) & 0.204(1) & 0.023(1) & 0.015(1) & 0.006(1) & 5.156(43) & 3.893(64) &
2.938(160) & -- \\
V31 & 15.078(1) & 0.252(1) & 0.042(1) & 0.023(1) & 0.018(1) & 4.589(24) & 2.822(43) &
1.658(54) & -- \\
V55 & 15.086(1) & 0.199(1) & 0.024(1) & 0.013(1) & 0.007(1) & 4.567(55) & 3.322(104)&
2.105(192) & --\\
V57 & 15.100(1) & 0.251(1) & 0.044(1) & 0.018(1) & 0.017(1) & 4.769(18) & 2.853(42) &
1.471(46) & -- \\
V60 & 15.115(1) & 0.252(1) & 0.045(1) & 0.022(1) & 0.017(1) & 4.639(24) & 2.683(49) &
1.570(60) & -- \\
V62 & 15.078(1) & 0.240(1) & 0.045(1) & 0.020(1) & 0.016(1) & 4.531(37) & 2.818(81) &
1.409(102) & -- \\
V78 & 15.117(1) & 0.197(1) & 0.029(1) & 0.007(1) & 0.006(1) & 4.636(30) & 2.680(128) &
1.464(154) & -- \\
V79 & 15.018(1) & 0.190(1) & 0.019(1) & 0.008(1) & 0.004(1) & 4.607(47) & 4.121(117) &
3.167(210) & -- \\
V80 & 15.095(1) & 0.201(1) & 0.013(1) & 0.015(1) & 0.006(1) & 4.979(92) & 3.870(79) &
2.480(196) & -- \\
V88 & 15.056(1) & 0.212(1) & 0.024(1) & 0.018(1) & 0.006(1) & 4.582(49) & 3.430(64) &
2.720(193) & -- \\
V95 & 15.050(1) & 0.245(1) & 0.042(1) & 0.020(1) & 0.020(1) & 4.763(28) & 2.900(60) &
1.532(60) & -- \\
V98 & 15.094(1) & 0.232(1) & 0.033(1) & 0.019(1) & 0.008(1) & 4.582(40) & 3.164(68) &
2.241(152) & -- \\
V100 & 15.146(1) & 0.252(1) & 0.040(1) & 0.023(1) & 0.016(1) & 4.688(34) & 2.906(62) &
1.771(85) & -- \\
V105 & 15.234(2) & 0.297(3) & 0.049(4) & 0.032(4) & 0.028(4) & 4.733(75) & 2.972(116)
& 1.788(136) & -- \\
V108 & 14.980(1) & 0.236(2) & 0.020(2) & 0.012(2) & 0.010(2) & 4.894(101) & 3.537(166)
& 2.715(192) & -- \\
V113 & 15.101(2) & 0.263(2) & 0.053(2) & 0.020(2) & 0.020(2) & 4.799(44) & 3.022(112)
& 1.338(114) & -- \\
V116 & 14.972(1) & 0.238(1) & 0.021(1) & 0.017(1) & 0.012(1) & 4.861(43) & 3.898(53) &
2.537(77) & -- \\
V117 & 14.951(1) & 0.169(2) & 0.018(2) & 0.011(2) & 0.001(2) & 5.344(114)& 3.601(192)&
3.233(355)& -- \\
V125 & 15.098(2) & 0.229(2) & 0.028(2) & 0.021(2) & 0.012(2) & 4.816(82) & 3.284(111)&
2.270(200) & -- \\
V128 & 15.109(1) & 0.243(2) & 0.034(2) & 0.025(2) & 0.015(2) & 4.635(46) & 3.358(61) &
2.360(103) & -- \\
V132 & 15.035(1) & 0.191(1) & 0.026(1) & 0.010(1) & 0.008(1) & 4.654(53) & 2.908(129)
& 1.255(296) & -- \\
V133 & 14.989(1) & 0.225(1) & 0.034(1) & 0.018(1) & 0.014(1) & 4.912(43) & 3.202(81) &
1.910(101) & -- \\
V139 & 14.796(1) & 0.185(1) & 0.026(1) & 0.017(1) & 0.010(1) & 4.705(43) & 2.950(65) &
1.610(110) & -- \\
V161 & 15.161(1) & 0.209(2) & 0.023(2) & 0.011(1) & 0.005(1) & 4.777(66) & 3.556(145)
& 2.652(296) & -- \\
\hline
\end{tabular}
\label{fourier_coeffs}
\end{table*}

These Fourier parameters and the semi-empirical calibrations of Jurcsik
\& Kov\'acs (1996), for RRab stars, and Morgan, Wahl \& Wieckhorts (2007),
for RRc stars, are used to obtain [Fe/H]$_{\rm ZW}$ on the Zinn \& West (1984) 
metallicity scale which have been transformed
to the UVES scale using the equation [Fe/H]$_{\rm UVES}$=
$-0.413$ +0.130[Fe/H]$_{\rm ZW}-0.356$[Fe/H]$_{\rm ZW}^2$ (Carretta et al. 2009).
The absolute magnitude $M_V$ can be derived from the calibrations of Kov\'acs \&
Walker
(2001) for RRab stars and of Kov\'acs (1998) for the RRc stars. The effective
temperature
$T_{\rm eff}$ was estimated using the calibration of Jurcsik (1998). For brevity we
do not explicitly present here the above mentioned calibrations; however, the
corresponding equations, and most
importantly their zero points, have been discussed in detail in previous papers
(e.g. Arellano Ferro et al. 2011; 2013) and the interested reader is referred to them.

\noindent 
Let us recall that the calibration for [Fe/H] for RRab stars of Jurcsik
\& Kov\'acs (1996) is applicable to RRab stars with a {\it deviation parameter}
$D_m$,
defined by Jurcsik \& Kov\'acs (1996) and Kov\'acs \& Kanbur (1998), not exceeding an
upper limit. These authors suggest $D_m \leq 3.0$. The $D_m$ is listed in
column 10 of Table~\ref{fourier_coeffs}. However we have relaxed the criterion
to include six stars with 3.0 $\leq D_m \leq$ 4.8.  The RRab V25 and V157, which are
likely blended in our images, are not included in the Fourier decomposition analysis.
They are addressed in $\S$  \ref{sec:IND_STARS}.

\begin{table*}
\footnotesize
\caption[] {\small Physical parameters for the RRab and RRc stars. The numbers in
parentheses indicate the uncertainty on the last 
decimal place.}
\centering
\label{fisicos}
\hspace{0.01cm}
 \begin{tabular}{lccccccc}
\hline 
Star&[Fe/H]$_{\rm ZW}$  &[Fe/H]$_{\rm UVES}$ &$M_V$ & log~$T_{\rm eff}$  & log$(L/{\rm
L_{\odot}})$ &
$M/{\rm M_{\odot}}$&$R/{\rm R_{\odot}}$\\
\hline
 &  &  & RRab stars &  & & \\
\hline
V3 &$-1.375(23)$& $-1.265(23)$& 0.610(3) & 3.805(8) & 1.656(1) & 0.64(6) & 5.55(1) \\
V5 &$-1.410(11)$& $-1.304(11)$& 0.584(1) & 3.813(7) & 1.666(1) & 0.68(6) & 5.40(1) \\
V6 &$-1.436(17)$& $-1.333(17)$& 0.603(3) & 3.812(8) & 1.659(1) & 0.68(6) & 5.39(1) \\
V7 &$-1.512(22)$& $-1.423(23)$& 0.658(3) & 3.816(8) & 1.637(1) & 0.71(7) & 5.16(1) \\
V9 &$-1.494(16)$& $-1.402(17)$& 0.452(1) & 3.797(8) & 1.719(1) & 0.67(6) & 6.18(1) \\
V11 &$-1.480(13)$& $-1.386(14)$& 0.507(1) & 3.808(7) & 1.697(1) & 0.70(6) & 5.73(1) \\
V12 &$-1.412(19)$& $-1.306(19)$& 0.658(3) & 3.821(8) & 1.637(1) & 0.72(6) & 5.05(1) \\
V16 &$-1.617(21)$& $-1.555(24)$& 0.410(4) & 3.802(8) & 1.736(2) & 0.74(7) & 6.16(1) \\
V17 &$-1.578(10)$& $-1.505(12)$& 0.482(1) & 3.807(7) & 1.707(1) & 0.72(6) & 5.83(1) \\
V20 &$-1.317(17)$& $-1.202(16)$& 0.547(2) & 3.807(8) & 1.681(1) & 0.65(6) & 5.66(1) \\
V30 &$-1.447(23)$& $-1.347(23)$& 0.603(3) & 3.806(8) & 1.659(1) & 0.65(6) & 5.55(1) \\
V32 &$-1.424(17)$& $-1.320(17)$& 0.682(3) & 3.821(7) & 1.627(1) & 0.73(6) & 4.99(1) \\
V33 &$-1.548(18)$& $-1.468(20)$& 0.641(3) & 3.815(8) & 1.643(1) & 0.71(6) & 5.21(1) \\
V34 &$-1.434(19)$& $-1.331(19)$& 0.613(2) & 3.808(8) & 1.655(1) & 0.66(6) & 5.45(1) \\
V38 &$-1.359(18)$& $-1.247(17)$& 0.675(3) & 3.819(8) & 1.630(1) & 0.72(6) & 5.04(1) \\
V39 &$-1.541(11)$& $-1.459(12)$& 0.512(1) & 3.808(7) & 1.695(1) & 0.71(6) & 5.71(1) \\
V41 &$-1.436(12)$& $-1.334(13)$& 0.608(1) & 3.820(7) & 1.657(1) & 0.73(6) & 5.19(1) \\
V54 &$-1.388(14)$& $-1.279(14)$& 0.657(3) & 3.822(7) & 1.637(1) & 0.74(6) & 5.01(1) \\
V59 &$-1.363(13)$& $-1.252(13)$& 0.623(1) & 3.813(7) & 1.651(1) & 0.66(6) & 5.31(1) \\
V61 &$-1.398(16)$& $-1.291(16)$& 0.596(1) & 3.810(7) & 1.662(1) & 0.66(6) & 5.47(1) \\
V64 &$-1.464(16)$& $-1.367(17)$& 0.626(1) & 3.811(7) & 1.649(1) & 0.67(6) & 5.35(1) \\
V74 &$-1.274(20)$& $-1.156(18)$& 0.666(3) & 3.824(8) & 1.633(1) & 0.72(7) & 4.95(1) \\
V77 &$-1.562(47)$& $-1.484(53)$& 0.315(2) & 3.786(11) & 1.774(1) & 0.68(9) & 6.93(1)
\\
V81 &$-1.426(13)$& $-1.322(13)$& 0.614(1) & 3.810(7) & 1.654(1) & 0.66(6) & 5.40(1) \\
V82 &$-1.428(12)$& $-1.324(13)$& 0.617(1) & 3.810(7) & 1.653(1) & 0.66(6) & 5.40(1) \\
V83 &$-1.381(21)$& $-1.272(21)$& 0.641(3) & 3.811(8) & 1.644(1) & 0.65(6) & 5.33(1) \\
V86 &$-1.529(16)$& $-1.443(18)$& 0.499(3) & 3.812(7) & 1.700(1) & 0.73(6) & 5.65(1) \\
V89 &$-1.401(11)$& $-1.294(11)$& 0.606(1) & 3.811(7) & 1.658(1) & 0.67(6) & 5.41(1) \\
V90 &$-1.452(13)$& $-1.352(14)$& 0.490(2) & 3.815(7) & 1.704(1) & 0.73(6) & 5.60(1) \\
V91 &$-1.389(15)$& $-1.281(15)$& 0.603(1) & 3.808(7) & 1.659(1) & 0.65(6) & 5.49(1) \\
V94 &$-1.418(31)$& $-1.313(32)$& 0.616(5) & 3.815(9) & 1.654(2) & 0.68(7) & 5.29(1) \\
V103 &$-1.437(27)$& $-1.335(28)$& 0.622(3) & 3.808(9) & 1.651(1) & 0.66(7) & 5.43(1)\\
V107 &$-1.587(20)$& $-1.516(22)$& 0.665(3) & 3.813(7) & 1.634(1) & 0.69(6) & 5.21(1)\\
V110 &$-1.124(27)$& $-1.009(22)$& 0.598(3) & 3.809(9) & 1.661(1) & 0.62(7) & 5.47(1)\\
V114 &$-1.529(21)$& $-1.444(23)$& 0.568(1) & 3.804(8) & 1.673(1) & 0.67(6) & 5.69(1)\\
V118 &$-1.424(20)$& $-1.321(20)$& 0.517(3) & 3.812(8) & 1.693(1) & 0.68(6) & 5.60(1)\\
V119 &$-1.458(21)$& $-1.359(22)$& 0.626(3) & 3.810(8) & 1.650(1) & 0.67(6) & 5.38(1)\\
V123 &$-1.344(22)$& $-1.231(21)$& 0.601(1) & 3.805(8) & 1.660(1) & 0.64(6) & 5.57(1)\\
\hline
Weighted &&&&&&\\
Mean & $-1.444(3)$& $-1.335(3)$ & 0.575(1) & 3.811(1) & 1.670(1) & 0.68(1) & 5.49(1)\\
\hline 
 &  &  & RRc stars &  & & \\
\hline
V15 &$-1.45(13) $& $-1.35(13) $& 0.553(2) & 3.865(0) & 1.679(1) & 0.51(0) & 4.32(1) \\
V31 &$-1.62(8) $ & $-1.56(09) $& 0.559(1) & 3.866(0) & 1.676(0) & 0.59(0) & 4.28(1) \\
V55 &$-1.67(20)$ & $-1.62(24)$ & 0.582(5) & 3.863(1) & 1.667(2) & 0.52(5) & 4.30(1)\\
V57 &$-1.43(7) $ & $-1.32(07) $& 0.571(1) & 3.870(0) & 1.672(0) & 0.60(0) & 4.19(1) \\
V60 &$-1.52(8) $ & $-1.43(09) $& 0.575(1) & 3.869(0) & 1.670(0) & 0.61(0) & 4.20(1) \\
V62 &$-1.41(14) $& $-1.30(14) $& 0.589(2) & 3.870(0) & 1.665(1) & 0.60(0) & 4.15(1) \\
V78 &$-1.27(21) $& $-1.15(19) $& 0.647(1) & 3.873(1) & 1.641(1) & 0.59(1) & 3.98(1) \\
V79 &$-1.26(24) $& $-1.15(22) $& 0.588(2) & 3.867(1) & 1.665(1) & 0.49(0) & 4.21(1) \\
V80 &$-1.46(16) $& $-1.36(17) $& 0.559(4) & 3.865(0) & 1.676(2) & 0.51(0) & 4.31(1) \\
V88 &$-1.61(12) $& $-1.55(14) $& 0.588(2) & 3.864(0) & 1.665(1) & 0.52(0) & 4.27(1) \\
V95 &$-1.48(10) $& $-1.38(11) $& 0.553(1) & 3.869(0) & 1.679(0) & 0.61(0) & 4.25(1) \\
V98 &$-1.52(12) $& $-1.43(13) $& 0.597(2) & 3.867(0) & 1.661(1) & 0.55(0) & 4.20(1) \\
V100 &$-1.51(11) $& $-1.43(12) $& 0.571(1) & 3.868(0) & 1.672(1) & 0.59(0) & 4.23(1)\\
V105 &$-1.49(20) $& $-1.39(22) $& 0.514(3) & 3.868(1) & 1.694(1) & 0.62(1) & 4.33(1)\\
V108 &$-1.56(32) $& $-1.48(36) $& 0.554(4) & 3.865(1) & 1.678(2) & 0.53(1) & 4.33(1)\\
V113 &$-1.33(20) $& $-1.22(19) $& 0.556(2) & 3.871(1) & 1.678(1) & 0.61(1) & 4.20(1)\\
V116 &$-1.56(11) $& $-1.48(12) $& 0.530(2) & 3.863(0) & 1.688(1) & 0.51(0) & 4.40(1)\\
V117 &$-1.60(38) $& $-1.54(43) $& 0.568(19)& 3.864(1) & 1.673(4) & 0.51(1) &4.32(2)\\
V125 &$-1.42(20) $& $-1.31(21) $& 0.575(4) & 3.868(1) & 1.670(1) & 0.56(1) & 4.21(1)\\
\hline
\end{tabular}
\end{table*}

\begin{table*}
\footnotesize
\addtocounter{table}{-1}
\caption{Continued}
\centering
 \hspace{0.01cm}
 \begin{tabular}{lccccccc}
\hline 
Star&[Fe/H]$_{\rm ZW}$  &[Fe/H]$_{\rm UVES}$ &$M_V$ & log~$T_{\rm eff}$  & log$(L/{\rm
L_{\odot}})$ &
$M/{\rm M_{\odot}}$&$R/{\rm R_{\odot}}$\\
\hline 
V128 &$-1.41(11) $& $-1.30(11) $& 0.566(2) & 3.868(0) & 1.674(1) & 0.56(0) & 4.23(1)
\\
V132 &$-1.39(22) $& $-1.28(22) $& 0.616(2) & 3.870(1) & 1.654(1) & 0.57(1) & 4.09(1)
\\
V133 &$-1.36(15) $& $-1.25(14) $& 0.568(2) & 3.870(0) & 1.673(1) & 0.58(0) & 4.20(1)
\\
V139 &$-1.56(12) $& $-1.48(13) $& 0.590(2) & 3.867(0) & 1.664(1) & 0.57(0) & 4.21(1)
\\
V161 &$-1.58(28) $& $-1.51(32) $& 0.578(3) & 3.864(1) & 1.669(1) & 0.51(1) & 4.29(1)
\\
\hline
Weighted&&&&&\\
Mean & $-1.49(3)$ & $-1.39(3)$ & 0.578(1) & 3.867(1) & 1.669(1) &0.56(1) & 4.22(1) \\
\hline
\hline
\end{tabular}
\end{table*}

The physical parameters for the RRL and the inverse-variance
weighted means are reported in Table~\ref{fisicos}.
Two independent estimations of [Fe/H]$_{\rm UVES}$ were found from 38 RRab and 24
RRc
stars in M5: $-1.34\pm0.11$ and $-1.39\pm0.12$ respectively, where the
uncertainties are the standard deviations of the inverse-variance
weighted means.

Morgan (2013) published a new calibration of 
[Fe/H]$_{\rm UVES}$ for RRc stars,
calculated for an extended number of stars and clusters, and noted that the
results of the new calibration are in general consistent with the old one from
Morgan et al. (2007), and that the
average differences are of the order of 0.02 dex. We have also carried out the
calculation
using the new calibration and found [Fe/H]$_{\rm UVES}=-1.43$, i.e. 0.04 dex
more
deficient than the value derived with the calibration of Morgan et al. (2007)
which is
in better agreement with the result from the RRab calibration. A similar conclusion
was found by Arellano Ferro et al. (2015b) for the case of NGC~6229.

The weighted mean $M_V$ values for the RRab and RRc stars are 0.575$\pm$0.082 mag
and
0.578$\pm$0.028 mag respectively and will be used in section
\ref{sec:distance} to estimate the mean distance to the parent cluster.

\section{The horizontal branch of M5}
\label{sec:HB}

\subsection{Evolved and binary RR Lyrae stars}

Recently we analysed in detail the HB structure of NGC 6229 (Arellano
Ferro et al. 2015b), a cluster that apparently has a similar history and
evolution to that of M5. We comment here on their similarities and some important
differences. From an analysis of the proper motion of M5,  Cudworth \& Hanson (1993),
Cudworth (1997) and Scholz et al. (1996) argued that M5, which is currently in the
inner-halo, has a wide galactic orbit reaching as far as 40-50 kpc from the
Galactic center, which reveals M5 to actually be, like NGC 6229, an outer-halo
cluster.
The similarity of metallicity and age between M5 and NGC 6229 has been noted in the
past (Borissova et al. 1999). In fact 
they have essentially the same metallicity; [Fe/H]$_{\rm UVES}$, -1.36 (M5) and -1.30
(NGC 6229) and their
CMDs indeed resemble each other, both having a long blue tail and a well populated
HB; their values of the Lee-Zinn parameter $\cal L$ ($\equiv (B-R)/(B+V+R)$ where $B,V,R$
refer to the number of stars to the blue, inside, and to the red of the IS)
are +0.31 and +0.24
respectively. 
Borissova et al. (1999) argued indetail in favour of the
two clusters being coeval to within 1 Gyr. NGC~6229 is an OoI cluster, as recently
discussed by Arellano Ferro et al. (2015b). In M5, the periods of the 
79 RRab and 34 RRc stars listed in Table \ref{variables} average 
0.553d and 0.311d, respectively; these values, along with the ratio of the number of
RRab to RRc stars, clearly confirm that M5 is also an
Oosterhoff type I cluster. Both clusters have a substantial number of SR
variables and are poor in SX Phe stars, since three SX Phe are known in M5 and only
one in NGC~6229.
However, when it comes to the instability strip (IS) in the HB, 
there are important differences: M5 has about twice as many stars in the HB 
as NGC 6629, which is a consequence of M5 being brighter and then more massive,
and hence it is not a surprise that it has about twice as many RRL. 
Nearly a hundred RRL in M5 show secular period variations (Szeidl et al. 2011) while
these have been detected
and measured in some 16  stars in NGC 6229 (Arellano Ferro et al. 2015b). We must
note however that, although the time base of available data is similar for M5 (at
least 100 years) and NGC 6229 ($\sim$83 years), the latter has not been observed as
heavily
as the former over the last century, which may have limited the discovery of secular
period changes.
A close comparison of the HB in M5 and NGC 6229 has led Borisova et al. (1999)
to conclude that the blue HB in NGC 6229 is brighter than in M5, although the red HB
matches very well.

\begin{figure} 
\includegraphics[width=9.cm,height=13.cm]{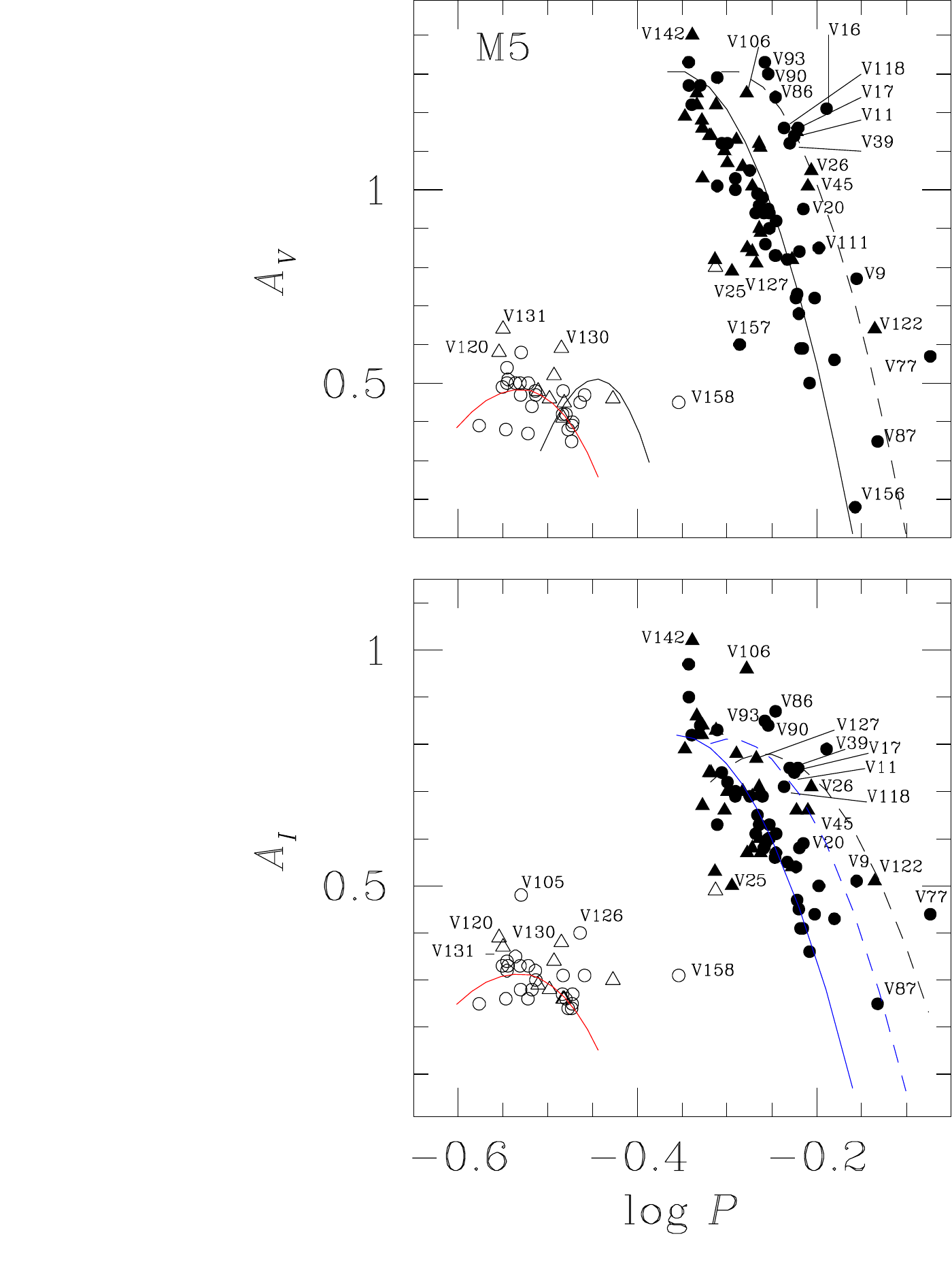}
\caption{The log P vs. amplitude plane for the $V$ and $I$
amplitudes of the RRL in M5.
Filled and open symbols represent RRab and RRc stars respectively. Triangles are used
for stars with Blazhko modulations. In the top panel the continuous and segmented
lines
are the loci found by Cacciari et al. (2005) for unevolved and evolved stars,
respectively, in the OoI cluster M3. In the bottom panel the black segmented locus
was found by Arellano Ferro et al. (2011, 2013) for the OoII
clusters NGC~5024 and NGC~6333. The blue loci are from
Kunder et al. (2013) for the OoI cluster NGC~2808. The red parabolas were
calculated by Arellano Ferro et al. (2015b) for a sample of RRc stars in five
OoI clusters. The black parabola was found by Kunder et al. (2013b) for 14 OoII
clusters.}
    \label{Bailey}
\end{figure}

The Bailey, or period-amplitude diagram, of M5 for the $V$ and $I$ filters is shown in
Fig. \ref{Bailey}. The scatter of the RRab stars about the general trends
represented
by the continuous lines derived for the OoI clusters M3 by Cacciari et al. (2005) in
$V$-band, and NGC 2808 by Kunder et al. (2013a) in $I$-band, is quite considerable and
a
number of outliers
sitting on the dashed sequence can be identified and they have been labelled in the
figure. According to Cacciari et al. (2005) this is the locus for the evolved stars.
The Bailey diagram further confirms M5 as an OoI type cluster. 

\begin{figure} 
\includegraphics[width=8.3cm,height=5.5cm]{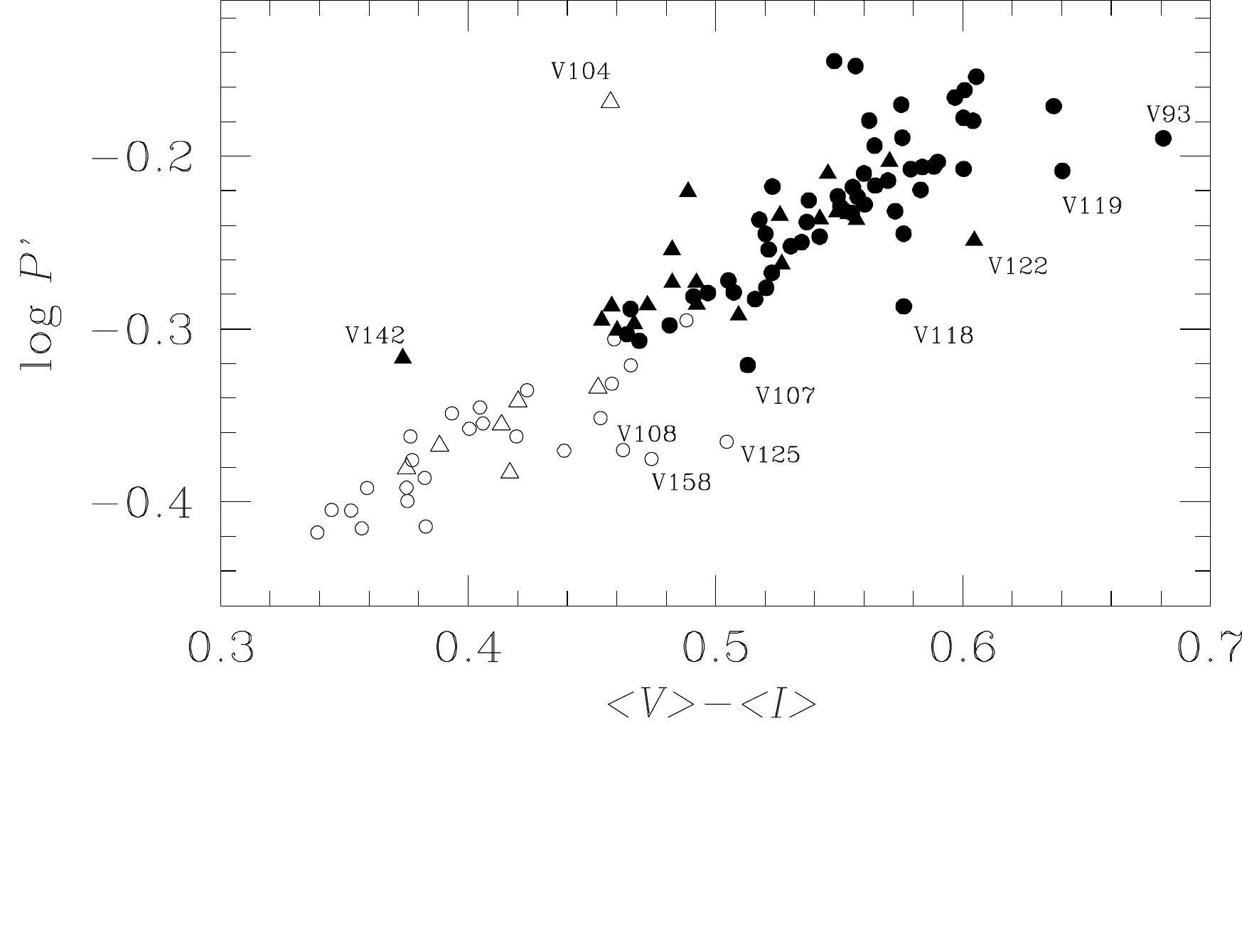}
\caption{Period-colour diagram of stable RRL stars. Filled circles are
used for RRab stars and empty circles for RRc stars. Triangles are used for stars
with Blazhko modulations. The periods of the
RRc stars were fundamentalized by adding 0.128 to log~$P$. The reduced
period $P'$ is defined as $\log P' = \log P + 0.336(< V >-15.021)$.}
    \label{LP_COL}
\end{figure}

Several features can be highlighted from the distribution of RRL in the HB from its
expanded version in the bottom panel of Fig. \ref{CMD}, in
combination with
the Bailey diagram of Fig. \ref{Bailey} and the period-colour diagram displayed in
Fig. \ref{LP_COL}, very useful for the task of
interpreting the distribution of RRL stars in the instability strip. In
this last diagram, the period
$\log P' = \log P + 0.336(<V>-15.021)$, where 15.021 mag is the average $<V>$ for all
the
stable
RRL, is plotted as a function of $<V>-<I>$. The use of $P'$ significantly reduces the
scatter in the diagram (van Albada \& Baker 1971; Bingham et al. 1984).

Stable RRab and RRc stars as well as those exhibiting Blazhko modulations have been
distinguished in these three figures.
Let us note first that all RRab stars falling in the evolved sequence of the top
panel of Fig. \ref{Bailey} have been plotted with red filled circles and red labels in
the bottom panel of Fig.\ref{CMD}, and they are $bonafide$ the brightest stars, as
expected for stars in an
advanced evolutionary state towards the AGB. However a higher luminosity can be
attained by
non-evolutionary means, like the presence of an unseen companion or the existence of
atmospheric enrichment of helium due to extra mixing at the RGB stage (Sweigart
1997). Let us note the
labelled outlier stars in  Fig. \ref{LP_COL} (V93, V107, V108, V118, V119, V122, V125,
V158) (other peculiar outliers V25, V104 and V142 are discussed below in section
\ref{sec:IND_STARS}). These stars are labelled in the bottom panel of Fig. \ref{CMD}
in black.
These stars have too short a period for their colour which implies that they have
larger gravity and should be of lower luminosity. If in spite of this they happen to
be among the most luminous stars in the IS, then they are good candidates to have a
companion.
This seems to be the case of V107, V122 and V158. Similar cases were
identified by 
Cacciari et al. (2005) among the RRL in M3 (V48, V58, and V146) and by Arellano Ferro
et al.
(2015b) in NGC~6229 (V14, V31, V54 and V55). Thus, these stars with shorter periods
compared to others of similar colour, could have enhanced helium in their atmospheres
or have become luminous due to the presence of an unseen companion.
The remainder of the stars with red symbols might be truly evolved stars.
Unfortunately none of this was
included in the secular period change analysis of Szeidl et al. (2011) perhaps
because most of them were only discovered after 1987 and the time-base of existing
data is not long enough. 

\subsection{The RRab-RRc segregation and the Oosterhoff type}

It has been noted and discussed in several of our recent papers that some clusters
show a neat splitting in the distribution of stable RRab and RRc stars in the CMD
with the border at $(V-I)_0 \sim 0.45-0.46$. In Fig. \ref{CMD} the corresponding
border lines for NGC 6229, NGC~5024 and NGC~4590
duly reddened are indicated by vertical black dashed lines. A detailed discussion of
this fact 
has been given by Arellano Ferro et al. (2015b) (their section 5). In brief, a clear
RRab-RRc splitting is distinguished in the OoII clusters 
NGC~288, NGC~1904, NGC~5024, NGC~5053, NGC~5466, NGC~6333 and NGC~7099, 
all with rather blue HBs ($\cal L >$ 0.4). It is also observed in 
the OoII cluster NGC~4590 which has a red HB ($\cal L$=0.17)(see CMD in Fig. 11 of
Kains et al. 2015). Among OoI
clusters, NGC~3201 does not
present the splitting (Arellano Ferro et al. 2014) while NGC 6229 does, very
clearly (Arellano Ferro et al. 2015b).

\begin{figure} 
\includegraphics[width=8.0cm,height=7.5cm]{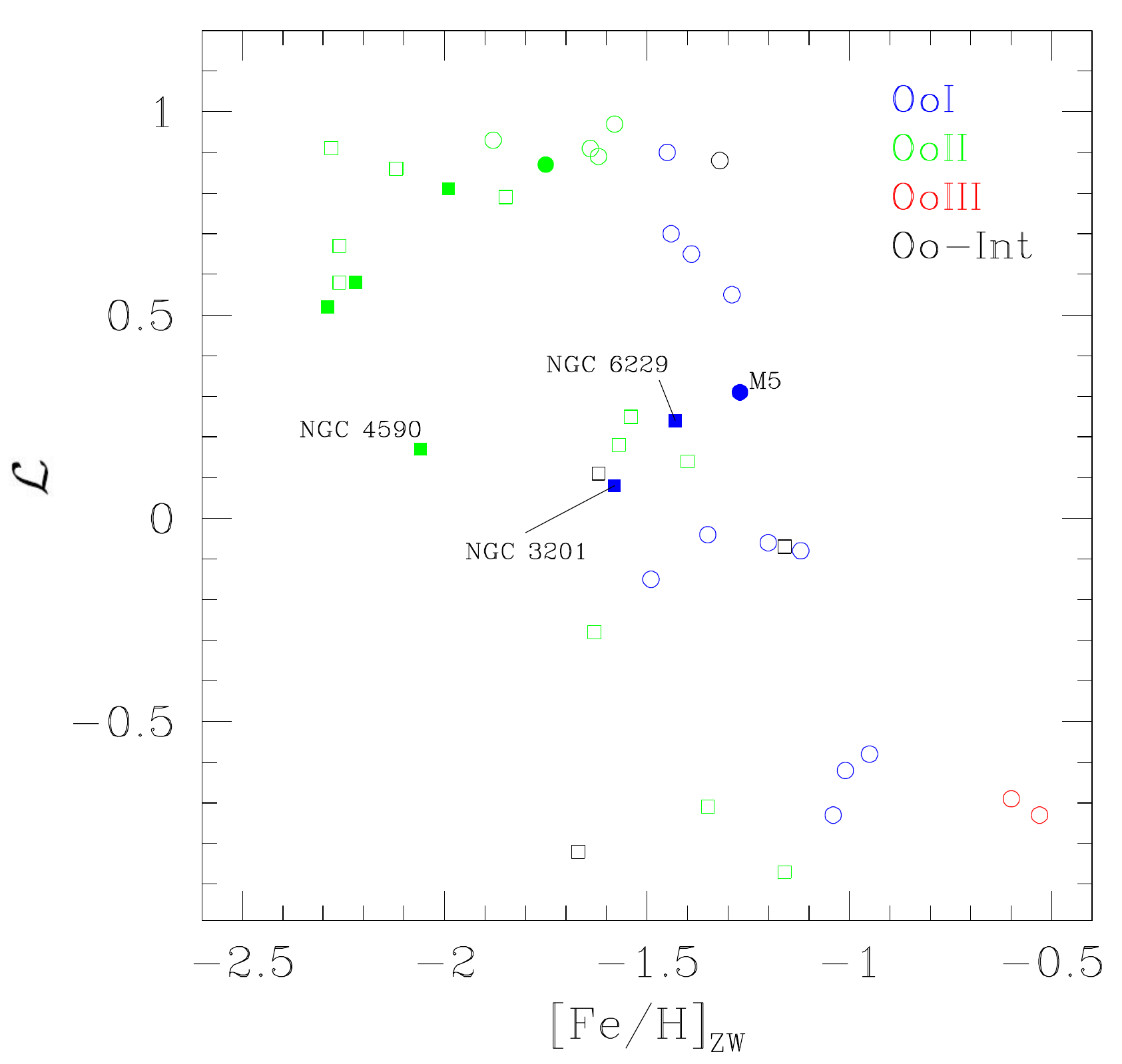}
\caption{Distribution of HB structure parametrized by the Lee-Zinn parameter $\cal L$
as a function of [Fe/H]$_{ZW}$ for Galactic GCs of different Oosterhoff
types, according to the legend, with data taken from Catelan (2009). Inner and
outer halo clusters are represented by circles and squares respectivelly}.
    \label{FeL}
\end{figure}

In Fig. \ref{FeL} we have plotted the Lee-Zinn parameter  $\cal L$ as a function of
[Fe/H]$_{ZW}$ taken mostly from Table 2 in Catelan (2009). The OoII clusters all show
blue tails, except NGC 4590. The more metallic OoI clusters may have both blue and red
HBs and display a large dispersion. Those clusters for which we have explored the RR
Lyrae distribution are plotted with filled symbols while empty symbols are used
otherwise. Three OoI clusters are also labelled; despite their closeness in the 
plot the only one with a clear RRab-RRc segregation is NGC 6229,
while NGC 3201, and now M5 (see Fig. \ref{CMD}), show both Blazhko and stable RRab
stars well distributed
all across the IS, in the either-or region, i.e. to the blue of the first-overtone red
border. It should be noted however that in all cases stable RRc stars are
well confined to the blue of the first-overtone red border as expected.
In passing let us mention that this is yet another important difference between M5
and "its twin" NGC 6229.

It is true that we have explored only three OoI clusters and that an analysis of a
larger number of both types of
Oosterhoff clusters is desirable; however, we might preliminarily conclude that blue-tailed OoII
clusters seem to have their RRab and RRc stars systematically well segregated around
the red border of the first-overtone instability strip. However, OoI clusters may or
may not have this property.

The distribution of RRL on the HB may be explained by the arguments of Caputo et al.
(1978) which involve the occurrence
of a hysteresis mechanism (van Albada \& Baker 1973) for stars crossing the IS.
According to this mechanism the stars in the  "either-or" region can retain the mode
they were pulsating in before entering the region, which depends on whether the star
is
coming from the blue side as an RRc or from the red side as an RRab. 
Caputo et al. (1978) also suggest that the original mode in which a RRL is pulsating
depends on the location of the starting point at the Zero Age Horizontal Branch
(ZAHB), which in turn depends on the mass and the chemical composition of the star;
they used these ideas to explain the existence of the two Oosterhoff groups as
follows; 1) the ZAHB point is in the fundamental zone, leading to an
assortment of RRc and RRab in the "either-or" region and a lower value of the
average period for both RRc and RRab and a lower proportion of RRc stars,
hence an OoI cluster, 
and 2) the ZAHB point is bluer than the fundamental zone, in the "either-or" or in the
first overtone regions, then the "either-or" region is populated exclusively by RRc
stars and the RRab stars are to be found only in the fundamental region, producing 
larger averages of periods and a higher proportion of RRc and hence an OoII type
cluster.

The above scenario explains well the clean segregation of RRc and RRab seen
in the OoII clusters, or the definitive lack of segregation in OoI
clusters like NGC~3201 or M5, but, as commented by Arellano Ferro et al. (2015b), the
clean segregation observed in the OoI cluster NGC~6229 is at odds with the
above picture.

Furthermore, RRab-RRc segregation in OoII clusters is also favoured by the arguments
of Pritzl et al. (2002) that in clusters with blue HBs, i.e. large values of $\cal L$,
stars with masses below a critical mass on the ZAHB to the blue of the IS, evolve
redwards and spend sufficient time to contribute to the population of RRL.
Thus, clusters with a blue HB morphology are OoII, with
redwards evolution, which in turn favours the segregation between RRc and RRab on the
IS which, in passing, should at least statistically display increasing secular
period variations. Unfortunately there are few clusters with a large enough number of
RRL with secularly changing periods that have been studied; e.g. the OoII
clusters Omega Cen (Martin 1938) and M15 (Silbermann \& Smith 1995) and the OoI
clusters M3 (Corwin \& Carney 2001) and M5 (Szeidl et al. 2011); in all these
clusters only a small surplus, and probably not statistically significant, of RRL are
reported as having increasing periods.

Growing evidence of the existence of multiple stellar populations in GCs (see for
instance Jang \& Lee (2015) and the references therein) invites us to consider
whether this is connected with the observed RRab-RRc splitting discussed above.
According to Jang \& Lee (2015), in inner-halo GCs the time elapsed between the first
stellar generation (G1) and a helium ($Y$) and CNO enhanced second generation (G2) is
$\sim$ 0.5 Gyr while in the outer-halo GC's, G1 has been delayed by $\sim$ 0.8 Gyr and
the
time between G1 and G2 has been longer, $\sim$ 1.4 Gyr. 
They have demonstrated that for metallicities [Fe/H]$\sim -1.5$, in outer GC's the IS
is mostly (but not exclusively) populated by G1 while the $Y$- and CNO-enhanced G2 RR
Lyrae stars must be more luminous (see their figure 7) with an average $<P_{ab}> =
0.552$d, which agrees very well with the
observed $<P_{ab}> = 0.553$d in M5. For inner GC's of similar metallicity the mix
of G1 and G2 in the IS is likely more due to the shorter time between generations in
which
case $<P_{ab}> = 0.607$d. For lower metallicity outer GC's, [Fe/H]$\sim -2.0$, the IS
is
populated basically by G2 stars. 
We shall recall that for a $Y$- and CNO-enhanced
G2, the masses on the ZAHB are shifted towards lower temperatures (Jang et al. 2014,
their figure 3), hence displacing the first overtone red-edge (FRE) to the red (Bono
Caputo \& Marconi 1995). More than one generation sharing the IS will have different
FRE boundaries contributing to a mix of RRc and RRab stars.

What effect the above scenario might have on the presence
or not of the RRab-RRc splitting is not clear at present as several factors seem 
to play a role; the overall metallicity, the presence of more than one
generation in the IS and the time elapsed between generations. We could add to this 
the possible presence of pre-ZAHB RR Lyrae stars (Silva Aguirre et al. 2008).
Clearly more investigation of a larger sample of GC's is necessary to confront the
observations with theoretical predictions.
We speculate that in outer-halo OoII clusters the larger
time difference between generations favours the existence of an RRab-RRc splitting as
we
have generally observed.

\section{Distance and metallicity of M5 from its variable stars}
\label{sec:distance}

The distance to M5 can be estimated from our data using different approaches; firstly,
from the weighted mean $M_V$, calculated for the RRab and RRc from the Fourier 
light curve decomposition (Table \ref{fisicos}),
which can be considered as independent estimates since they come from different
empirical
calibrations and zero points. Secondly, we can use the $I$-band RR Lyrae P-L
relation derived by Catelan, Pritzl \& Smith (2004). As a third approach we can use
the three known SX Phe stars and their P-L relation. And finally a fourth approach
is via the bolometric magnitude for stars at the tip of the
RGB. Below we expand on these solutions.
 
Being M5 a nearby cluster, its reddening is small, we adopted
$E(B-V)$=0.03 mag (Harris 1996). Given the mean $M_V$ for RRL in Table \ref{fisicos}
we
found a true distance modulus of 
$14.403\pm0.067$ mag and $14.387\pm0.094 $ mag using 38 RRab and 24 RRc stars
respectively, which correspond to the distances 
$7.6 \pm 0.2$ and $7.5 \pm 0.3$ kpc.
The quoted uncertainties are the standard deviations of the corresponding means.
The distance to M5 listed in the catalogue of Harris (1996) (2010 edition) is
7.5 kpc in good agreement with our calculations.

The $I$-band RR Lyrae P-L relation derived by Catelan, Pritzl \& Smith (2004) is
of the form:

 \begin{equation}
M_I = 0.471 - 1.132 {\rm log} P + 0.205 {\rm log} Z,
\label{eqn:PL_RRI}
\end{equation}
 
\noindent
with ${\rm log} Z =[M/H] -1.765$. We applied these equations to all 62 RRab and
RRc stars in 
Table \ref{fisicos}. The periods for the RRc stars were fundamentalized following the
period ratio $P_1/P_0 =0.7454$ in double mode stars (Catelan 2009). We found an
average distance of 7.2$\pm$0.3 kpc.

\begin{figure} 
\includegraphics[width=7.5cm,height=6.0cm]{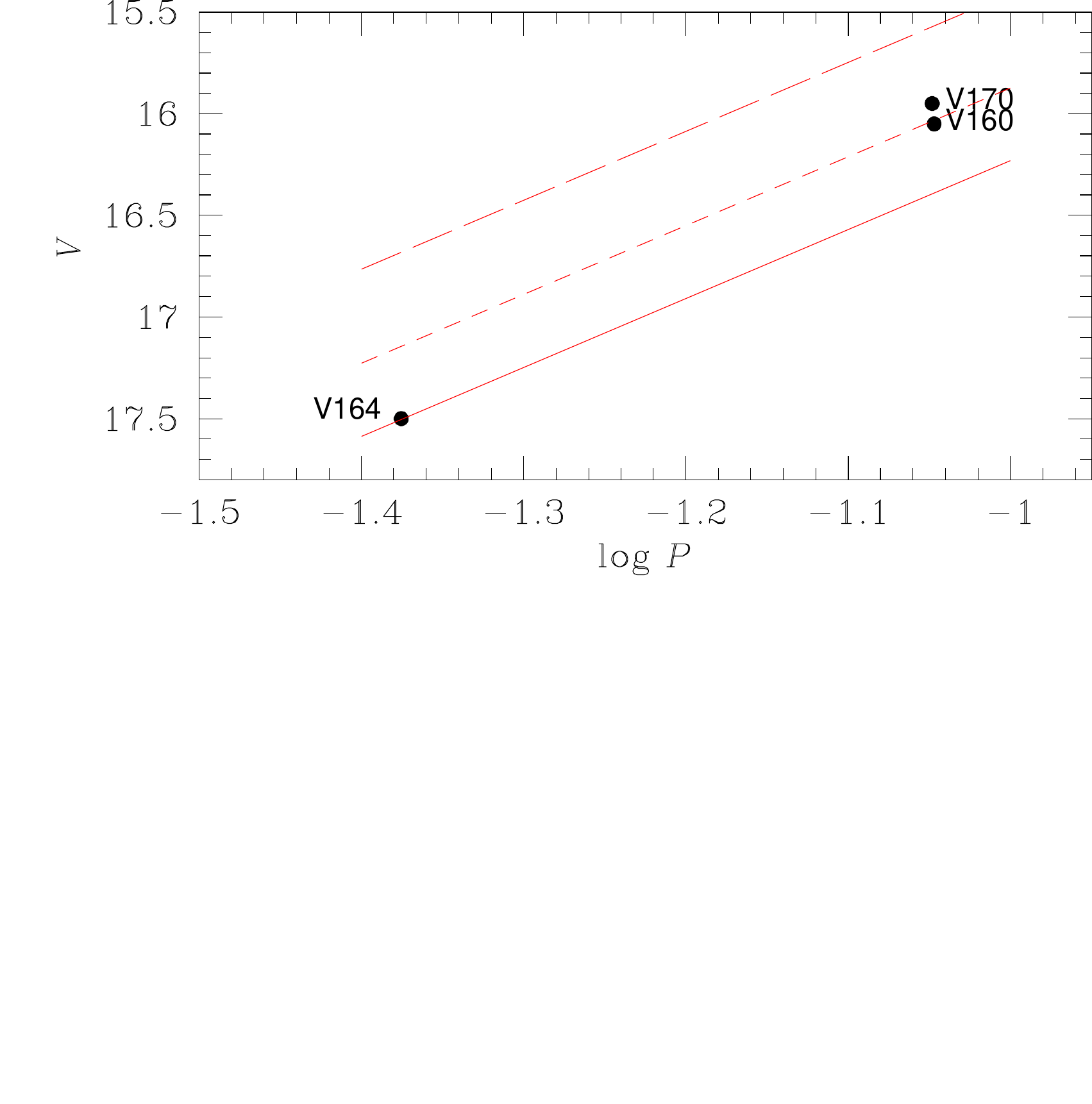}
\caption{M5 SX phe stars in the Period-Luminosity plane. The lines are the P-L
relationship for the SX Phe fundamental mode(continuous) of  Cohen \& Sarajedini
(2012), first-overtone (short segmented) and second-overtone (long
segmented).}
    \label{SXPL}
\end{figure}

An independent estimate of the distance can be obtained from the three SX Phe known in
the cluster. Fig. \ref{SXPL} shows the P-L relationship of Cohen \& Sarajedini
(2012);

 \begin{equation}
M_V = -(1.640 \pm 0.110) - (3.389 \pm 0.090) {\rm log} P_f
\label{eqn:PL_CS}
\end{equation}

\noindent
positioned for the RRL mean distance. The first-overtone and second-overtone lines
were positioned adopting the frequency ratios $F/1O$=
0.783 and $F/2O$ = 0.571 (see Santolamazza et al. 2001 or Jeon et al.
2003; Poretti et al. 2005). It seems clear from the plot that V164 is a fundamental
pulsator while V160 and V170 pulsate in the first-overtone. Then, adopting eq.
\ref{eqn:PL_CS} we find an average
true distance modulus of $14.428 \pm 0.150$ or a distance of
$7.7 \pm 0.4$ kpc. The uncertainty was calculated from calibration errors neglecting
the uncertainty in the period.

Yet another approach to the cluster distance determination is by using the tip of
the
RGB. This method was developed for estimating distances to nearby galaxies (Lee,
Freedman \& Madore, 1993). One possibility is to use the specific calibration of
the bolometric magnitude of the tip of the RGB, of M5 by Viaux et al.
(2013) under the arguments that
the neutrino magnetic dipole moment enhances the plasma decay process,
postpones helium ignition in low-mass stars, and therefore extends the red giant
branch (RGB) in GCs. These authors came to the conclusion that in
M5 $M_{bol}^{tip}$= -4.17 $\pm$ 0.13 mag and we adopted that value.
As in Arellano Ferro et al. (2015b) for the case of NGC 6229, we found that the method
is extremely sensitive to the star selection. Reasonable results are found for the
stars nearest to the tip of the RGB, i.e. the reddest and brightest in this
region of the CMD.  We restricted our calculation
to the two brightest RGB stars in our sample V50 and V174. According to Viaux et
al. (2013), the TRGB is between 0.05 and 0.16 mag brighter than the brightest stars
on the RGB. The fact that V50 and
V174 may not be the brightest stars on the RGB implies that their magntiudes would
have to be corrected to bring them to the TRGB by at least the above
quantities; the larger the correction the smaller the resulting distance. We
applied a correction between 0.05 and 0.16 mag to V50 and V174 and found a mean
distance of between 7.5 and 7.2 kpc respectively.

The above independent estimations of the cluster distance are, within their
uncertainties, satisfactorily in agreement. However, given the sensitivity of the SX
Phe P-L and the TRGB approaches to the number of stars involved as well as to the
sample selection, 
we believe that the distance determinations from the RRL stars, which deal with much
larger samples and carefully calibrated zero points of the luminosity scale are
the best achieved.

For the metallicity, the overall average value from the RRL is [Fe/H]$_{\rm
UVES}=-1.36 \pm 0.16$ or [Fe/H]$_{\rm ZW}=-1.47 \pm 0.14$ found from the Fourier
decomposition of the light curves of 38 RRab and 24 RRc. These values can be compared
with previous values in the literature:
[Fe/H]$_{\rm ZW} = -1.29$ (Harris 1996); [Fe/H]$_{\rm ZW} = -1.23$ (Kaluzny et al.
2000) from an independent Fourier decomposition analysis; [Fe/H]$_{\rm ZW} = -1.44$
(Zinn 1985); [Fe/H]$_{\rm UVES} = -1.11 \pm 0.03$ (Carretta et al. 2009)

\subsection{On the age of M5}
\label{sec:age}

The age of M5 has been discussed in numerous works. We have not attempted an
independent estimation. The age of M5 was determined by 
Jimenez \& Padoan (1998) via its luminosity function and they found an age of  
$10.6\pm0.8$ Gyr assuming [$\alpha$/Fe]=+0.4. 
Two more recent determinations of the age of M5 are by Dotter et al. (2010) using
relative ages from isochrone fitting, and by VandenBerg et al. (2013) using
an improved calibration of the ”vertical method” or the magnitude
difference between the turn off point and the HB; these authors
find 12.25$\pm$0.75 Gyr and 11.50$\pm$0.25 Gyr respectively. Thus, in the
CMD of Fig. 3 we have overlayed the corresponding isochrones for
12.0 Gyr from the Victoria-Regina evolutionary models (VandenBerg et al. 2014) and
for the metallicities [Fe/H]=$-1.42$ (red) and
[Fe/H]=$-1.31$ (blue) and [$\alpha$/Fe]=+0.4, shifted for the average apparent
distance modulus $\mu$=14.395 mag found from the RRab and
RRc stars, and $E(V-I) = 1.616 E(B-V) = 0.048$.

The isochrones for these two metallicities are very similar except
perhaps at the tip of the RGB where the isochrone for the more metal-rich
composition gets less luminous. However, if shifts are applied to these
isochrones within the uncertainty of the distance modulus and reddening, $\sim 0.1$
mag, 
the two cases are indistinguishable. We can stress then
that our CMD is consistent with the metallicity and distance derived
in this paper and an age of $12.0\pm1.0$ Gyr found in the above
papers for M5.

\section{Summary and discussion}
\label{summary}
The Fourier decomposition of the light curves of stable RRL, and the
calibrations and zero points available in the recent literature allowed us to
determine
the mean metallicity and distance to M5 as [Fe/H]$_{\rm UVES}=-1.34\pm0.11$ and
$-1.39\pm0.12$ and $7.6\pm 0.2$ kpc and $7.5\pm 0.3$ kpc, from the RRab and RRc
stars respectively. Also individual values of radius and mass are provided.
 The employment of the $I-band$ RR Lyrae P-L relation leads to a distance of
7.2$\pm$0.3 kpc.
The distance to the cluster was also estimated from two independent methods; the P-L
relation of SX Phe and the luminosity at the tip of the RGB finding 7.7$\pm$0.4
kpc and 7.5$\pm$0.1 kpc respectively. However we have shown that in the case of
M5 the accuracy of these two alternative methods does not compete with the accuracy
attained with the RR Lyraes and the Fourier light curve decomposition.

A group of 16 evolved stars have been identified from their distribution in the
amplitude-period plane. Evolved stars should be moving to
the red in the CMD, hence their periods should be increasing. From the period
change
study of Szeidl et al. (2011) and our own period change analysis, which will be
published elsewhere, we can see that some stars have increasing periods
(V11, V16, V39, V45, V77, V87 and V90) and some have clearly decreasing periods (V9,
V12
and V17), which might be 
unexpected in truly evolved stars. However it is understood now that stochastic 
processes may produce both positive, negative or irregular period variation, e.g.
mixing events in the core of a star at the HB (Sweigart \& Renzini 1979) or, as
suggested more recently by Silva Aguirre et al. (2008), period decreases can occur in
pre-ZAHB RR Lyrae stars on their final approach to the ZAHB.
In the bottom panel of Fig.
\ref{CMD} we have drawn a small line segment from the corresponding circles, to the
right or left to indicate period increase or decrease respectively. The remainder of
the evolved stars were not found to be undergoing secular period variations in our
analysis. From their position on
the CMD, their long period value and large period change rate, we conclude that 
V77 and V87 are truly advanced in their evolution towards the AGB.

For the most luminous stars in the IS (V107, V122 and V158) we argue that they
are
not truly evolved stars but owe their overluminosity to a helium enhanced atmosphere
or to the presence of an unseen companion.

We note that in the HB instability strip of M5, RRab stars share the 
either-or region with RRc stars, like in the case of the
OoI cluster NGC 3201, but in contrast with NGC 6229, a cluster presumably nearly a
twin of M5, where the RRab-RRc splitting is noticeable. We also highlight the
fact that the RRab-RRc splitting is a common feature in OoII clusters with a blue
tail.

Our CCD $VI$ time-series, spanning a little more than two years, allowed us to
detect amplitude and phase modulations in 14 RRab and 9 RRc stars not previously
reported as having the Blazhko effect. These new findings account for incidence rates
of at least 38\% and 26\% of the RRab and RRc respectively showing the Blazhko effect
in M5.
This comes as a natural result since in the last few years the detection of
Blazhko variables in GCs has been increasing, most likely due to the
improvement in the quality of the CCD observations and reduction techniques. Take for
example the cluster NGC 5024 with the largest known population of Blazhko variables
with 66\% and 37\% of RRab and RRc respectively (Arellano Ferro et al. 2012).

Finally, it is well known from key studies in the past (e.g. Oosterhoff 1941, Szeidl
et al. 2011) that a large fraction of the RRL population of M5 is undergoing secular
period variations. Our present data add up to 20 years to the time-base of available
data for M5 and a re-analysis of the period change rates is deferred to a forthcoming
paper.

\section*{Acknowledgments}
To the memory of Janusz Kaluzny whose numerous contributions to the field of
variable stars in globular clusters have been a permanent source of inspiration.
Numerous suggestions and comments from an anonymous referee have been very valuable
and are
warmly acknowledged. It is a pleasure to thank No\'e Kains for his comments and
suggestions. This publication was made possible by grant
IN106615-17 from the DGAPA-UNAM (Mexico), the collaborative program CONACyT-Mincyt
(Mexico-Argentina) \# 188769 and by NPRP grant \#
X-019-1-006 from the National Research Fund (a member of Qatar Foundation).
We have made an extensive use of the SIMBAD and ADS services, for which we
are thankful.

\appendix

\section{Comments on individual stars and the Blazhko variables}
\label{sec:IND_STARS}

In this section we comment on the light curves, variable types and nature of some 
interesting or peculiar variables in Table \ref{variables} and Figs. \ref{VARSabA} and
\ref{VARSc}. We put some emphasis
on the amplitude and phase modulations of the Blazhko type  
in specific stars. In all the stars
labelled '$Bl$' in Table \ref{variables} the amplitude variations are neatly
distinguished in
the light curves in Figs. \ref{VARSabA} and \ref{VARSc} in both the $V$ and $I$
filters.

{\bf V25, V36, V53, V74, V102, V108, V140 and V159}. All these stars
were found by Arellano Ferro et al. (2015a) to be misidentified in the literature. In
that paper the identifications have been discussed and corrected and a detailed
identification chart is given. 

{\bf V14}. In the CVSGC (2014 update) it is noted as an unconfirmed variable,
probably after Evstigneeva et al. (1995). However the star is clearly
variable in the study of Kaluzny et al. (2000) and in the present work. It displays
some amplitude modulation already noted by Kaluzny et al. (2000).

{\bf V18}. It displays a tremendous amplitude variation from 0.617 mag in 2012 to
1.218 mag in 2013 and 2014. The observations on January 23, 2013 are already
consistent with the large amplitude, meaning that the star underwent the amplitude
change between May 2012 and January 2013. The photographic light curve from 1934
from Oosterhoff
(1941), although a bit scattered, shows an amplitude of 0.76 mag and while not
directly comparable with our $V$ or $I$ light curve amplitudes we mention that 
Oosterhoff himsef noted considerable light curve variations and suggested to
classifiy it as an irregular variable. The data from 1997
of Kaluzny et al. (2000) show an amplitude of 1.27 mag and a mild suggestion of the
Blazhko effect. The repetitivity of our light curves from February to May in 2012  at
the
low amplitude and again during 2013-2014 at the large amplitude, suggests that the
star remains with
a constant amplitude before it goes through the amplitude variation episodes: it is 
therefore a very
strong Blazhko modulator with a rather long period. According to Szeidl et al.
(2011)
the Blazhko period is longer than 500d.

{\bf V25}. This star is heavily blended with a nearby star which explains the
relatively noisy light curve in Fig. \ref{VARSabA}. It has been discussed
and clearly
identified by Arellano Ferro et al. (2015a).

{\bf V27}. Despite its relative isolation it shows a very peculiar light curve. We
have not been able to identify more than one period. Thus the observed
amplitude and phase modulations must be due to the Blazhko effect. We adopted the
period
of Szeidl et al. (2011) that phases the light curve best. They did not find secular
period changes for this star. The Blazhko modulations are very prominent.

{\bf V28}. Similar to V18, this star shows a very pronounced amplitude
modulation from 1.120 mag to 0.653 mag with much different time distributions than
V18. The amplitudes in the light curves of Oosterhoff (1941) and Kaluzny et al.
(2000) are in between the above two extremes.

\begin{figure*}
 \centering
\includegraphics[width=15.0cm,height=16.cm]{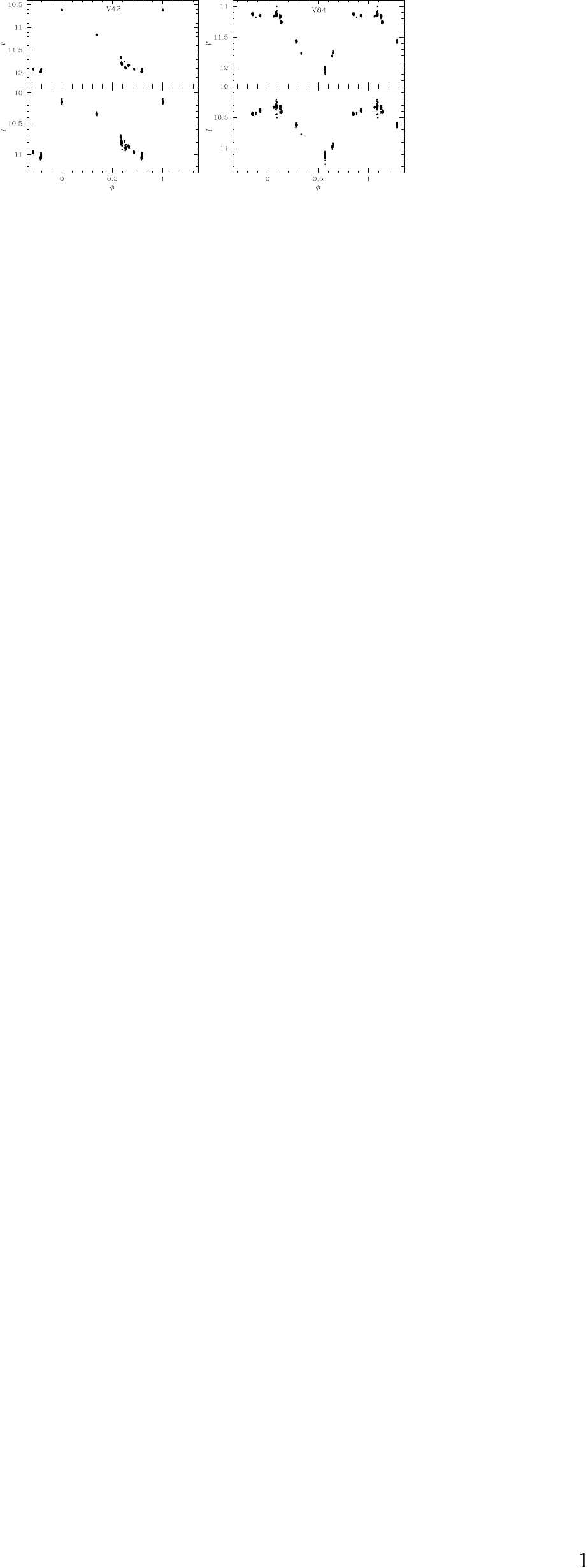}
\caption{Light variations of the CW stars V42 and V84 phased with the ephemerides
given in Table \ref{variables}.}
\label{CWs}
\end{figure*}

{\bf V42, V84}. These two W Virginis (CW) stars are shown in the CMD of Fig. \ref{CMD}
and their
light curves are  displayed in Fig. \ref{CWs}. The data are included in
the electronic Table \ref{tab:vi_phot}. Their periods are listed in Table
\ref{variables}. For V42 the period 25.735d given in the CVSGC phases our data
correctly. For V84 the period 53.95d in the CVSGC does not phase our data properly but
we found that about half of it, 26.49d, produces a nice light curve.

{\bf V50,V171-V181.} A discussion and the light curves of all these semi-regular
late-type (SRA) variables can be found in the paper by Arellano Ferro et al. (2015a).

{\bf V77}. This is the RRab with the largest period in M5 (0.845158d). It has a low
amplitude and a peculiar shape of the light curve with a roundish maximum. In the
amplitude-period diagram
(Fig. \ref{Bailey}) the star falls in the extreme right of the diagram, much beyond
the
evolved stars sequence. In the CMD (bottom of Fig. \ref{CMD}) it is one of the 
brightest stars. The star is undergoing a secular period increase with a rate
among the largest in M5 (Szeidl et al. 2011).
All these properties are consistent with the star being in an advanced evolutionary
state towards the AGB.

{\bf V78}. This star has the shortest period among the known RRL stars in M5
(0.264820d). Kaluzny et
al. (2000) have argued that it may be a second overtone pulsator (RRe) and noted the
slight asymmetry of the light curve. We point out that while the asymmetry is there,
it appears
as prominent as in the other RRc stars with longer periods (e.g. V15, V55, V60 etc.).
In
fact, an inspection of the light curves of the RRc stars in Fig. \ref{VARSc} reveals
that
a slight asymmetry is present in most of them, so this seems to be more the rule than
the exception in this cluster.  However,  following Catelan's (2004) practice, we
have checked the log P - $<I>$ plane
and confirmed that, after fundamentalizing the period of V78 assuming that it is a
second-overtone pulsator, the star falls in the natural extension to short period of
the RRab distribution. This seem to support the idea that V78 is a RRe.

{\bf V87}. Like V77, this star has a long period and a small amplitude and it is also
an evolved star with the corresponding large period increase rate (Szeidl et al.
2011).

{\bf V101}. A discussion of our data of this cataclysmic variable of the U~Gem
type and its light curve is given by Arellano Ferro et al. (2015a).

{\bf V104}. It has been noticed by several authors that the light curve of this
peculiar star shows very prominent amplitude and phase modulations. It has been
suggested to be a double mode star (Reid 1996) and a semi-detached binary (Drissen \&
Shara 1998). Its rather long period for an RRc star and its
position among RRab stars on the log P - $<I>$ plane make it more likely to be an RRab
star. We used {\tt period04} to search for traces of double mode activity but
found no significant secondary frequencies. Our light curve is very peculiar indeed
and shows multiple RR
Lyrae-like curves, all well phased with a period of 0.486748, with phase and mean
brightness displacements. It is probably an RRab in a binary system,
unfortunatelly we could only estimate five times of maximum on our data and two in
the light curve of Drissen \& Shara (1998) and thus, with the handful of times of
maximum the corresponding O-C diagram does not shed clear signs of duplicity.

{\bf V121}. This star is close to the center of the cluster but it is not blended
with another star of similar magnitude or brighter. Despite this, it shows
large
modulations in phase and amplitude probably due to the presence of Blazhko
effect. The light curves in Caputo el al. (1999) and
Drissen \& Shara (1998) look stable but their time-base are very short.

{\bf V127}. This star is very close to another of similar brightness. It is
identified in the discovering paper (Kravtsov 1988) and we confirm that the variable
is the NW of the pair (see Fig. \ref{chart}). In the paper by Olech et
al. (1999) its light curve appears as that of a small amplitude RRc-like with a period
0.544965d. However our period 0.540366d
produces a clear RRab light curve with amplitude apparently diminished due to flux
contamination by the neighbour. On the Bailey diagram (Fig. \ref{Bailey}) the star
clearly sits among the RRab stars. It also shows amplitude modulations.

{\bf V155}. A light curve of this W Ursae Majoris-type eclipsing binary (EW) is
presented and discussed in the paper by Arellano Ferro et. al. (2015a).

\begin{figure}
 \centering
\includegraphics[width=8.0cm,height=10.cm]{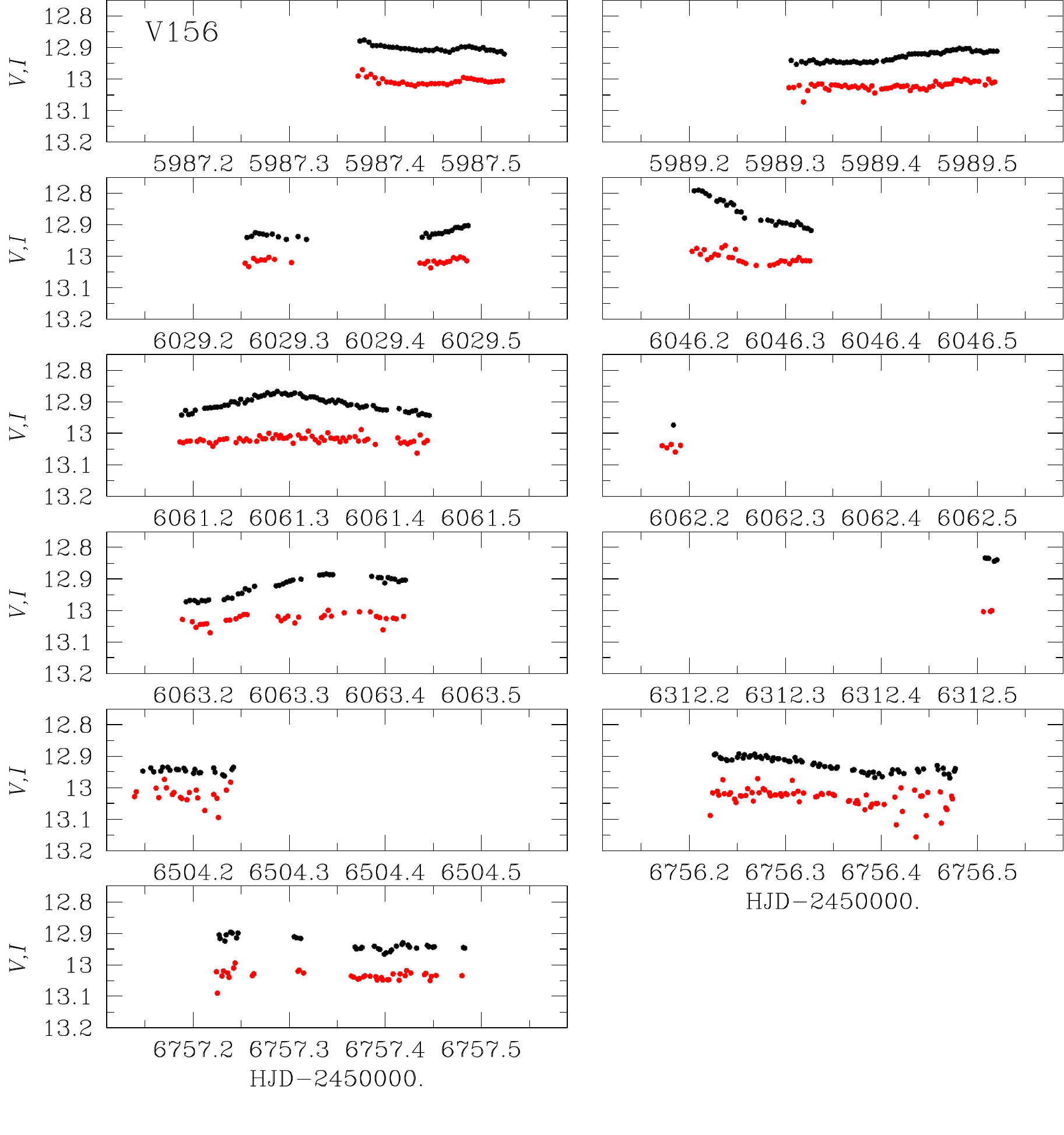}
\caption{Light variations of the RRab star V156. Black and red circles are the $V$ and
$I$ magnitudes, respectively. The $I$ magnitudes have been arbitrarily shifted to
accommodate them in the same box, so that the variations in both filters can be
appreciated.}
\label{V156}
\end{figure}

\begin{figure*}
 \centering
\includegraphics[width=11.5cm,height=23.cm]{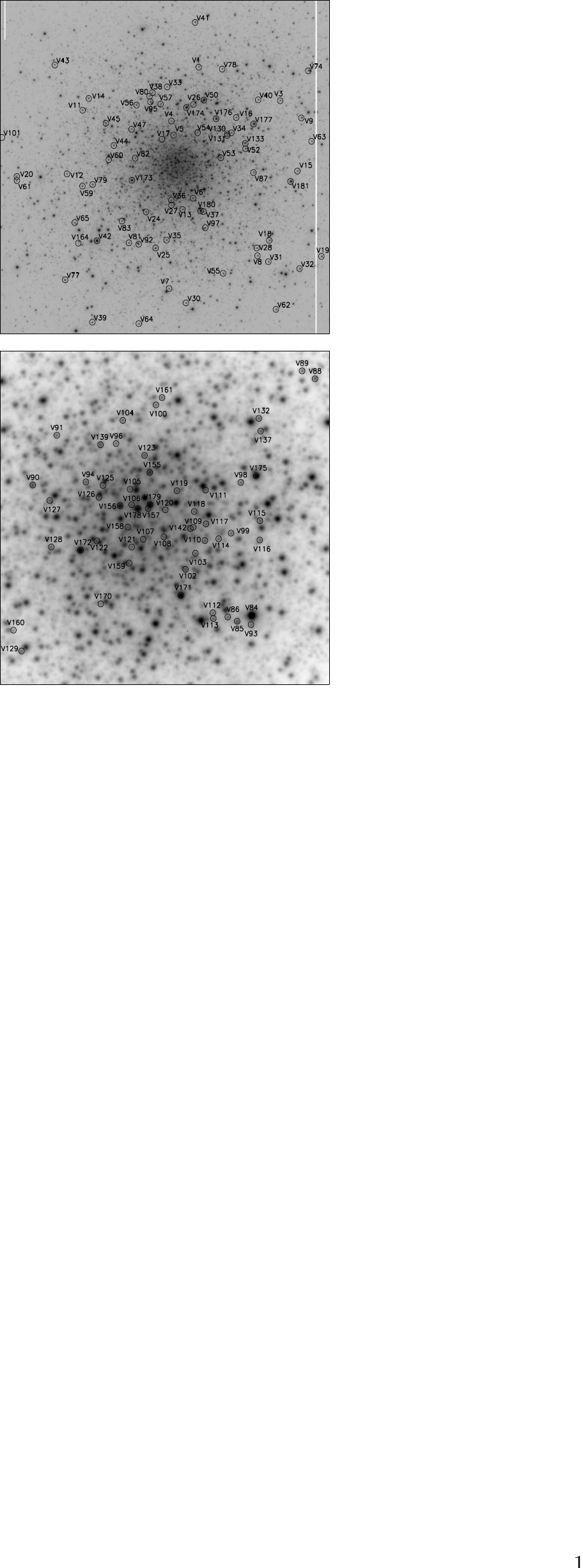}
\caption{Finding charts constructed from our $V$ reference image; north is up
and east is to the right. The cluster image at the top is
8.8$\times$8.8~arcmin$^{2}$. The core image at the bottom is
2.2$\times$2.2~arcmin$^{2}$. All of the variable stars listed in Table 3 are identified.}
\label{chart}
\end{figure*}

{\bf V156}. The variability of this star and a nice looking RRab light curve
were reported by Drissen \& Shara (1998) from their high
resolution images of the Hubble
Space Telescope (HST), and they labelled it as V15. The star identified by
these authors is strongly blended in our images and we were not
able to either calculate a period at confidence or recover a RRL-like
light curve, hence it was not included in Fig. \ref{VARSabA}. In Fig.
\ref{V156} we show the nightly light variations and note the changes of the
maximum brightness, inviting the suggestion of Blazhko modulations. Caputo et
al. (1999) offered an alternative identification and named the
star V156, however the light curve of that star does not show variations in our
collection.

{\bf V157}. The light curve of this star has an anomalously low amplitude, likely due
to
blending in our images with a close bright star. Its position in the CMD (Fig.
\ref{CMD}) and Bailey diagram 
(Fig. \ref{Bailey}) is therefore peculiar.

{\bf V158}. It is noted in the CVSGC as "RRc?" Indeed its period 0.442627d is rather 
long for an RRc, however, its light curve is a little asymmetric and resembles that of
the other RRc stars in the cluster.
In the Bailey's diagram (Fig. \ref{Bailey}), the star falls halfway between the RRc
and the RRab distributions. We have labelled it as RRc.

{\bf V159}. An eclipse for this eclipsing binary (E) was detected in our data.
The light curve is presented and discussed in the paper by Arellano Ferro et al.
(2015a).

{\bf V170}. The phased light curve for this SX Phe stars is given by Arellano
Ferro et al. (2015a).

\vspace{5 mm}
{\bf Interfering close pairs.}
We have identified a few close pairs of variable stars in our images that, depending
on the seeing conditions may or may not have contaminated each other's light
curves.
We
list them below and briefly discuss them.

{\bf V84-V93} The RRab V93 is 12.5 pixels or 3.7 arcsec from the bright CW V84 which
has produced the noisy appearance of the light curve in Fig. \ref{variables}.

{\bf V109-V142}. These two RRab stars are separated by only 4.0 pixels or 1.2 arcsec
in our images and hence the interference of each other's light curves which is evident
as
noise (see Table \ref{tab:observations}).
The light curve of V142 shows very prominent phase and amplitude
modulations. The variability was discovered by Brocato, Castellani \& Ripepi (1996)
but the light curve was not made available. Partial light curves were displayed
by Caputo et al. (1999) and Drissen \& Shara (1998) where the variations are clear but
no modulation was detected, most likely due to the limited time-base of these data
sets.

{\bf V112-V113}. The separation of these stars is 7.4 pixels or 2.2 arcsec which
means that in nights of poorer seeing their light curves are contaminated by the
light 
of the neighbour producing noisier light curves. 

{\bf V130-V131}. These stars are separated by 8.4 pixels or 2.5 arcsec, i.e.
ocassionally, on nights of poorer seeing there is 
mutual light contamination. Scatter in the light curve of V130  was noted by Kaluzny
et al. (2000) but their search for a secondary frequency
was futile. Our light curve
shows clear amplitude modulations which could be attributed to the Blazhko effect.
V131 is not included by
Kaluzny et al. Both stars included by Caputo et al. (1999) show no amplitude
modulations but their light curves are sparse. In our data V131 also shows mild
amplitude modulations. It is possible that the observed amplitude
modulations are due, at least partially, to the light contamination from the 
neighbour.

\vspace{5 mm}
{\bf RRab stars with Blazhko modulations}. A systematic and thorough analysis of 50
RRab stars in M5 was conducted by Jurcsik et al. (2011) in search of
Blazhko-type variability and its characterization, using a collection of photometry
of M5 covering nearly 100 years. They identified 20 Blazhko stars among the RRab
population and calculated the modulation periods (their Table 1). The modulation
periods range mostly between 40 and 600 days, with the exception of V72 for which the
modulation is about 1200 days.
The time-base of our data is 770 days (see Table \ref{tab:observations}). This means
that based only on our observations we should be able to detect the modulations in
all those variables, provided that the modulations are strong enough and the
photometry is of good quality. Of the 20 stars found by Jurcsik et al. (2011) four
(V2, V29, V58 and V72) are not in the FoV of our images. The remaining 16 stars are
labelled '$Bl^{b}$' in Table \ref{variables}, however V5, V30 and
V38, look rather stable in our light curves, thus we would
have been unable to identify these as Blazhko variables based exclusively on our data.
We find it difficult to judge the real Blazhko nature of these three stars. We
note
first that the Blazhko modulations reported by Jurcsik et al. (2011) are in all
cases very mild and that for V5 and V38 the Blazhko classification relies on a
comparison between the data from Reid (1996), whose light curves are a bit scattered,
and the data from Kaluzny et al. (1999; 2000) separated by 5 years; both stars are in
crowded fields and Kaluzny et al. 2000 has pointed out that neighbours at 5-10 arcsec
affect some of the photometric data. For the more isolated star V30 the comparison was
between the data of Storm et al. (1991) and Kaluzny et al. (1999; 2000) separated by
10 years,
reduced and transformed to the standard system by different approaches. While these
details may contribute to apparent amplitude differences, it is also possible that our
data time-base is not long enough for the detection of a long-period modulation.
For the other 13 stars the modulations are also evident in our data.

Besides the above mentioned stars, we have also been able to detect amplitude-phase
modulations in the light curves of another 14 RRab stars, some of which are subtle but
visible, hence probably
undetectable in old photographic data. These 14 newly identified Blazhko RRab stars
are labelled '$Bl$' in Table \ref{variables}. This makes a total of 30 
Blazhko stars in M5 in a sample of 79 RRab stars, i.e. an occurrence rate of
at least 38\%.

\vspace{5 mm}
{\bf RRc stars with Blazhko modulations}. No Blazkho RRc variables were reported by
Jurcsik et al.
(2011). However, the modulations are clearly visible in 9 stars of the 34 in
our sample; 
V35, V40, V44, V53, V55, V99, V120, V130 and V131. This makes an incidence rate
of at least 26\% for RRc Blazhko variables in M5.

\vspace{5 mm}
{\bf RRc stars with prominent bump near maximum}. The presence of bumps near the
maximum light in RRc stars is clear in V31, V35, V57, V60, V62, V95, V100, V105,
V108, V113, V116, V130 and V133. Bumps were noted before by Kaluzny et al. (2000) in
V31,
V35, V57 and V62 and highlighted that bumps do not occur for stars with periods above
0.28-0.30 d, which is true also for our extended list of "bumpy" stars with the
exception of V108 and V130 with P$\sim$0.328, although we may concede that the bump
in these cases is rather mild. Kaluzny et al. (2000) venture the
possibility that the bump is associated with some sort of interaction between
the first and second overtones. The bump was actually reproduced in the RRc
star U Com by hydrodynamical models constraining the metallicity and calibrating the
adopted turbulent convection model. The bump was found to be sensitive to the
parameters
governing the coupling between convection and pulsation (Bono, Castellani \& Marconi
2000).

\begin{figure*}
 \centering
\includegraphics[width=16.cm,height=23.cm]{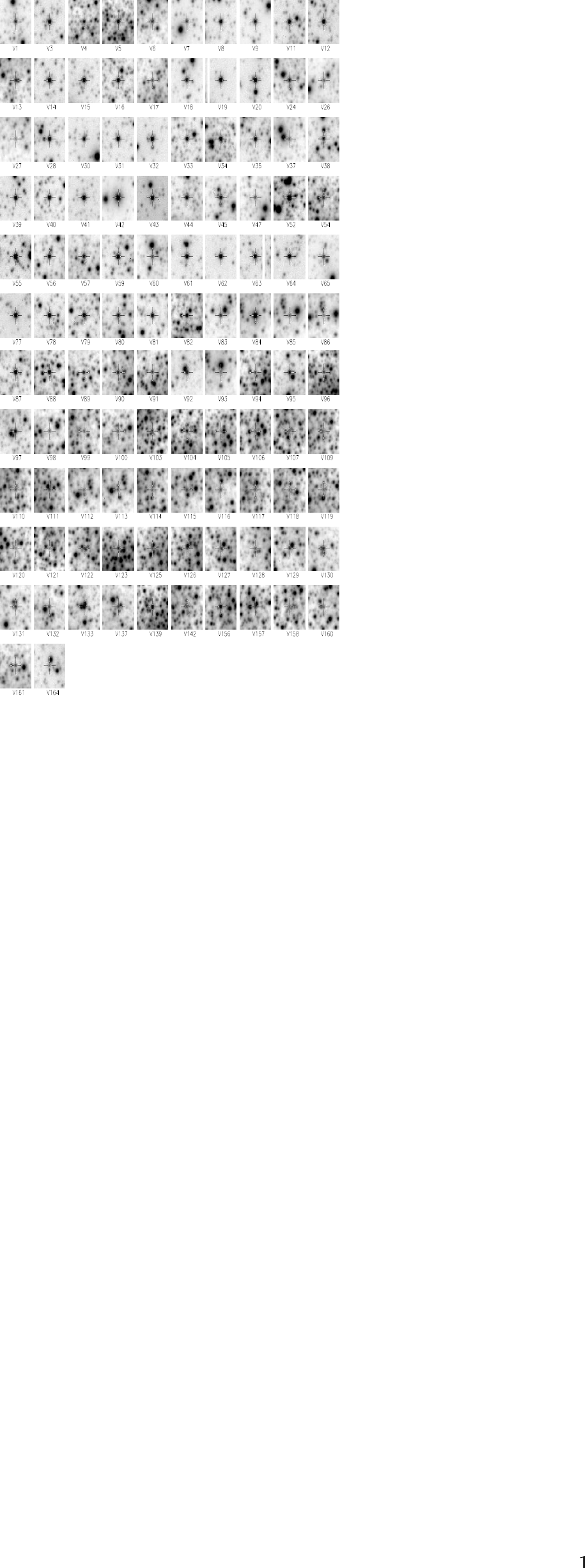}
\caption{Individual star cut-outs for each star in Fig. \ref{chart}. Images are of
size
23.7$\times$23.7~arcsec$^{2}$. All of the stars listed in Table 3, except for those with cutouts already published in Arellano Ferro et al. (2015a), are shown.}
\label{stamps}
\end{figure*}

\end{document}